\pdfoutput=1
\documentclass[aps,prd,amsmath,floats,floatfix, twocolumn,
superscriptaddress,nofootinbib,showpacs,longbibliography]{revtex4-1}
\usepackage[T1]{fontenc}
\usepackage[utf8]{inputenc}
\usepackage{txfonts}
\usepackage{verbatim}
\usepackage[dvipsnames, usenames]{xcolor}
\definecolor{linkcolor}{rgb}{0.0,0.3,0.5}
\usepackage[hypertexnames=false, unicode, colorlinks=true, linkcolor=linkcolor,
citecolor=linkcolor, filecolor=linkcolor,urlcolor=linkcolor,
pdfusetitle]{hyperref}
\usepackage[all]{hypcap}
\usepackage{graphicx}
\usepackage{xspace}
\usepackage{amssymb}
\usepackage{microtype}
\usepackage{subfigure}

\usepackage{url}

\usepackage[english]{babel}
\usepackage{blindtext}

\usepackage[normalem]{ulem} 
\usepackage{bm} 
\usepackage{mathrsfs}
\usepackage{xspace} 
\usepackage{fontawesome}
\usepackage[normalem]{ulem}

\DeclareMathAlphabet{\mathpzc}{OT1}{pzc}{m}{it}

\usepackage[nomessages]{fp}

\newcommand{\sk}[1]{}

\newcommand{\orcid}[1]{\href{https://orcid.org/#1}{\includegraphics[width=10pt]{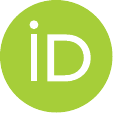}}}
\newcommand{\gwModel}{\textcolor{linkcolor}{\texttt{gwModel}}}
\newcommand{\gwModelX}{\textcolor{linkcolor}{\texttt{gwModel\_kick\_q200}}}
\newcommand{\gwModelXP}{\textcolor{linkcolor}{\texttt{gwModel\_prec\_flow}}}
\newcommand{\HLZ}{\textcolor{linkcolor}{\texttt{HLZ}}}
\newcommand{\NRSur}{\textcolor{linkcolor}{\texttt{NRSur7dq4Remnant}}}

\begin{document}
\title{Recoil kicks from binary black hole mergers in GWTC catalogs: implications for retention and hierarchical mergers}

\author{Tousif Islam\,\orcid{0000-0002-3434-0084}}
\email{tousifislam@ucsb.edu}
\affiliation{Kavli Institute for Theoretical Physics,\\University of California Santa Barbara, Kohn Hall, Lagoon Rd, Santa Barbara, CA 93106}

\hypersetup{pdfauthor={Islam et al.}}

\date{\today}

\begin{abstract}
We infer recoil (kick) velocities for both individual binary black hole (BBH) mergers and candidate intermediate-mass black hole events, as well as for the BBH populations inferred from GWTC catalogs up to GWTC–5. We obtain informative recoil constraints for several events, including GW231028\_153006 ($v_{\rm kick}=839^{+1018}_{-681} \,\mathrm{km\,s^{-1}}$) and GW231123\_135430 ($v_{\rm kick}=974^{+944}_{-760} \,\mathrm{km\,s^{-1}}$), while finding that the majority of event-level recoil posteriors remain broad and only weakly informative. We further infer consistent population-level recoil distributions across GWTC–3, GWTC–4, and GWTC–5, with median kick velocities of approximately $300$--$330\,\mathrm{km\,s^{-1}}$. Using both event-level and population-level recoil estimates, we find typical retention probabilities of $\sim2$--$3\%$ for globular clusters, $\sim28$--$32\%$ for nuclear star clusters, $\sim25$--$29\%$ for dwarf galaxies, and $\sim92$--$94\%$ for elliptical galaxies. We also compute recoil-induced displacements and dynamical-friction return times, finding that retained remnants in globular clusters and nuclear star clusters can remain displaced from their host cores for extended periods. Our results show that retention alone is not sufficient to determine the prospects for hierarchical mergers: hierarchical-merger efficiency depends on both remnant retention and post-kick re-centering.
\end{abstract}

\maketitle

\section{Introduction}
One of the most interesting aspects of binary black hole (BBH) mergers is that, due to linear momentum conservation, the remnant black hole receives a recoil kick~\cite{Maggiore:2007ulw,Maggiore:2007ulw}. Over the years, recoil velocities have been studied using both numerical and analytical frameworks, including numerical relativity (NR)~\cite{Baker:2006vn, Baker:2007gi, Baker:2008md,Herrmann:2006cd,Lousto:2007db,Herrmann:2007ac,Herrmann:2007ex,Herrmann:2007cwl,Holley-Bockelmann:2007hmm,Jaramillo:2011re,Koppitz:2007ev,Lousto:2008dn,Lousto:2010xk,Schnittman:2007ij,Sopuerta:2006et,Pollney:2007ss,Rezzolla:2010df,Lousto:2011kp,Lousto:2012gt,Lousto:2012su,Miller:2008en,Tichy:2007hk,Zlochower:2010sn,Healy:2014yta,Lousto:2009mf}, point-particle black hole perturbation theory (BHPT)~\cite{Nakano:2010kv,Sundararajan:2010sr,Islam:2023mob,Hughes:2004ck,Price:2013paa,Price:2011fm}, and post-Newtonian (PN)~\cite{Blanchet:2005rj,Sopuerta:2006wj,Sopuerta:2006et,Favata:2004wz,Fitchett:1983qzq,Fitchett:1984qn,Wiseman:1992dv,Kidder:1995zr} approximations. 

These studies have shown that the recoil is a highly nonlinear phenomenon. In particular, PN approximations, which are valid primarily during the inspiral phase of the binary evolution, are unable to accurately predict the kicks obtained from NR simulations, which solve the full nonlinear Einstein equations without relying on perturbative assumptions.
NR simulations have demonstrated that recoil velocities can be extremely large, reaching values of up to $5000\,\mathrm{km\,s^{-1}}$ or higher, especially for nearly equal-mass binaries with large in-plane spins~\cite{Campanelli:2007cga,Bruegmann:2007bri, Campanelli:2007ew, Choi:2007eu,Dain:2008ck,Gonzalez:2006md,Gonzalez:2007hi,Healy:2008js,Healy:2022jbh}. Consequently, substantial effort has been devoted to modeling the kick velocity across the binary parameter space by combining results from NR and BHPT calculations, guided by PN intuition~\cite{Zlochower:2015wga,Baker:2008md,Lousto:2008dn,Lousto:2010xk,Lousto:2012gt,Lousto:2012su,vanMeter:2010md,Healy:2014yta,Sundararajan:2010sr,Islam:2023mob,Varma:2019csw,Varma:2018aht,Merritt:2004xa,Kidder:1995zr,Sperhake:2019wwo,Islam:2025drw}. 
Current state-of-the-art models for recoil kick velocities include NR surrogate models such as \textcolor{linkcolor}{\texttt{NRSur7dq4Remnant}}~\cite{Varma:2019csw} and \textcolor{linkcolor}{\texttt{NRSur3dq8Remnant}}~\cite{Varma:2018aht}, which are developed for precessing-spin and aligned-spin binaries, respectively. In addition, semi-analytical models are available, including \HLZ{}~\cite{Lousto:2008dn,Lousto:2010xk,Lousto:2012gt,Lousto:2012su,Gonzalez:2007hi} for precessing-spin binaries and \gwModelX{}~\cite{Islam:2025drw} for aligned-spin systems. More recently, a normalizing-flow-based model, \gwModelXP{}~\cite{Islam:2025drw}, has also been developed for precessing-spin binaries.

\begin{figure*}
    \centering
    \includegraphics[width=\textwidth]{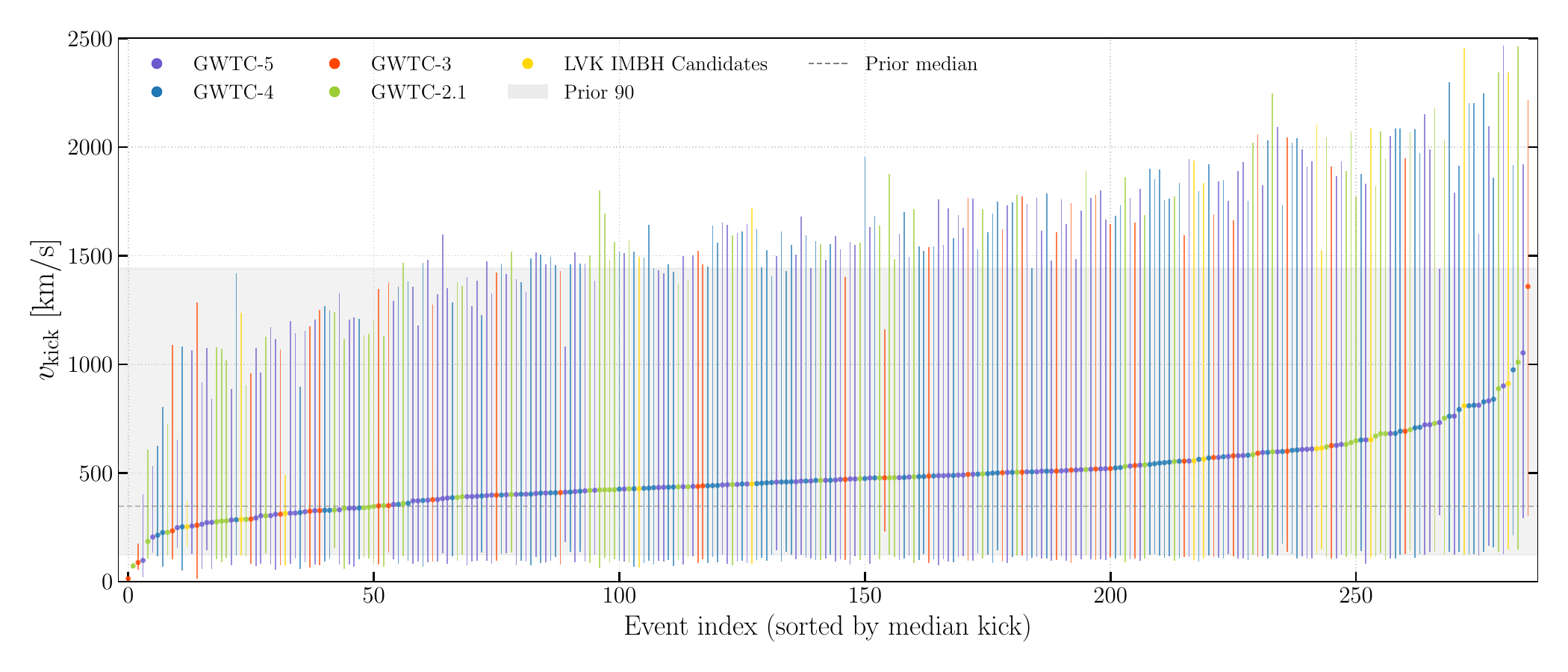}
    \caption{\textit{Recoil kicks inferred from individual GW events}. We show the median and 90\% credible intervals of the recoil kick velocity inferred from publicly available posterior samples for events in GWTC--5 (purple), GWTC--4 (blue), GWTC--3 (orange-red), GWTC--2.1 (green), and LVK candidate IMBH mergers (yellow). For comparison, the median and 90\% credible interval of the kick prior are shown as a horizontal dashed line and a gray shaded region, respectively. The events are ordered by their median inferred recoil kick velocities. While a small number of events exhibit noticeable deviations from the prior expectation, the majority of event-level recoil posteriors remain broad. More details are in Section~\ref{sec:event_kick}.}
    \label{fig:errorbar_kicks}
\end{figure*}

\begin{figure*}
    \centering
    \includegraphics[width=\textwidth]{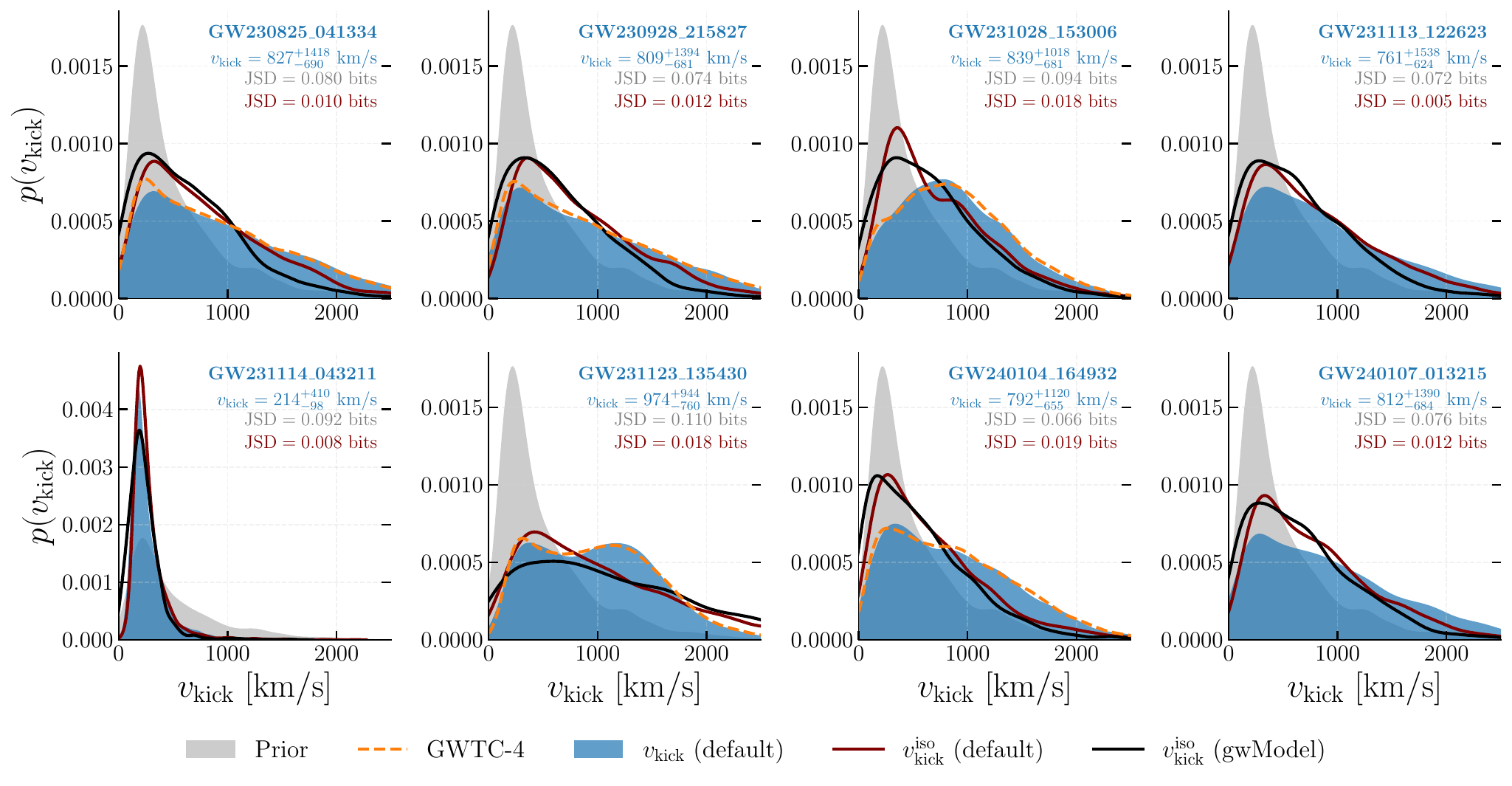}
    \includegraphics[width=\textwidth]{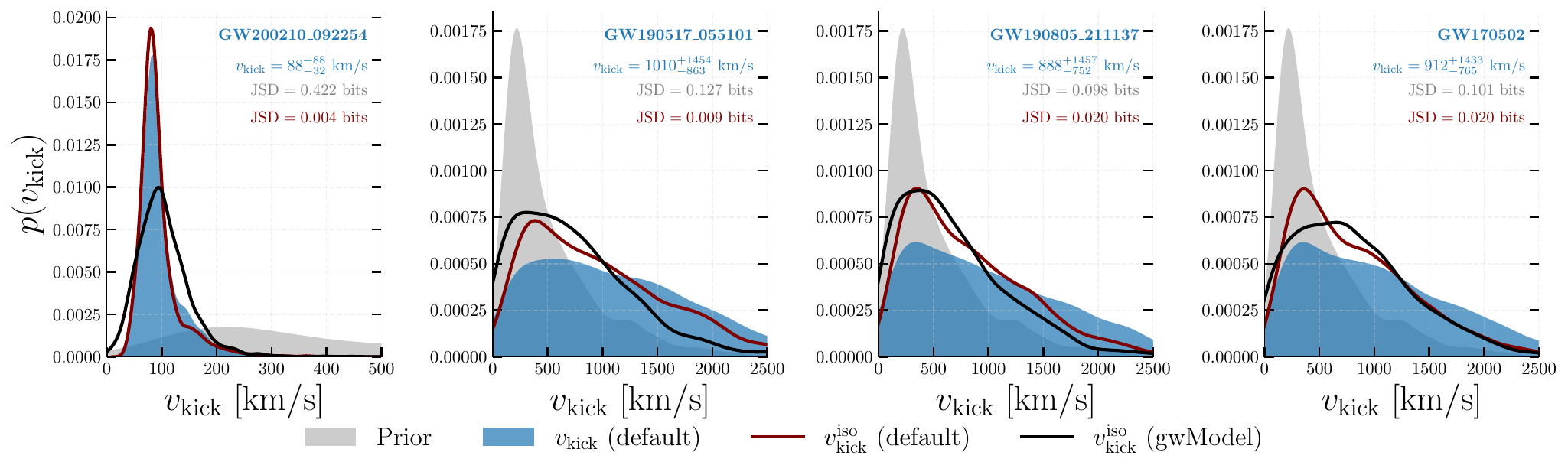}
    \caption{\textit{Example of events with informative kick measurements}. \textbf{Upper panel:} We show recoil kick velocity posteriors (shaded blue histograms) inferred from the publicly available GWTC--4 posterior samples for a subset of representative events. For comparison, we also show kick posteriors obtained by propagating only the mass ratio $q$ and spin magnitudes $|\chi_{1,2}|$ posteriors through the recoil prescription, while drawing the spin orientation angles isotropically (maroon histograms). The corresponding kick posteriors obtained using the \gwModel{} recoil prescription (for isotropic spin angles) are shown as black histograms whereas orange dashed histogram indicates the posteriors obtained from GWTC-4 data release. The JSD values between the default kick posterior and the prior are indicated in gray text, while the JSD values between the default posterior and the isotropic-angle case are indicated in maroon text. \textbf{Lower panel:} Same for selected events in GWTC--2.1 and GWTC--3 that have not already been analyzed in Ref.~\cite{Islam:2023zzj,Mahapatra:2021hme}. More details are in Section~\ref{sec:event_kick}.}
    \label{fig:select_gwtc4_kicks}
\end{figure*}

Astrophysically, recoil kicks have several important consequences~\cite{Merritt:2004xa,Gerosa:2016vip,Borchers:2025sid}. One of the most striking implications in the context of gravitational-wave (GW) sources is whether the remnant black hole is retained in its host environment, such as a globular cluster, an active galactic nucleus (AGN), or an elliptical galaxy. Under a simplified assumption, if the recoil velocity exceeds the escape velocity of the host environment, the remnant is likely to be ejected. Conversely, if the remnant is retained, it may subsequently merge with other black holes. This process, commonly referred to as a hierarchical merger, is thought to be a key channel for the formation of massive black holes in dense stellar systems~\cite{Holley-Bockelmann:2007hmm,Berti:2012zp,Gultekin:2004pm,Gerosa:2016vip,Borchers:2025sid,Gerosa:2021hsc,Gerosa:2017kvu}. 

Inferring recoil kicks from GW observations has therefore been an important area of study. One proposed approach involves searching for Doppler shifts imprinted on the waveform by the recoil of the remnant black hole~\cite{Gerosa:2016vip}. While this effect is theoretically measurable, in practice most detected events have low to moderate signal-to-noise ratios in current-generation detectors, making such signatures difficult to resolve.
An alternative strategy is to infer the binary's source parameters from the observed signal and then use these parameters to predict the resulting recoil velocity of the remnant~\cite{Varma:2020nbm}. This prescription, based on mappings calibrated to NR simulations, has been applied to estimate recoil kicks for several GW events (such as in Refs.~\cite{Varma:2022pld,Islam:2023zzj,Mahapatra:2021hme,CalderonBustillo:2022ldv}).

For example, using the latter methodology in conjunction with the \NRSur{} model, Ref.~\cite{Varma:2022pld} reported evidence for a large recoil kick in GW200129, with an inferred velocity of $1542^{+747}_{-1098}\,\mathrm{km\,s^{-1}}$. Ref.~\cite{Islam:2023zzj} subsequently analyzed a subset of 47 events from the GWTC-3 catalog and inferred a recoil velocity of $485^{+668}_{-252}\,\mathrm{km\,s^{-1}}$ for GW191109 using a similar prescription. They further identified two additional events for which informative kick measurements were obtained.
Ref.~\cite{Mahapatra:2021hme} employed the \HLZ{} framework to infer recoil velocities for the remnant black holes of 42 events from the GWTC-2 catalog. Most notably, they reported a well-constrained kick of $74^{+7}_{-10}\,\mathrm{km\,s^{-1}}$ for GW190814. Ref.~\cite{CalderonBustillo:2022ldv} computed the recoil for GW190412 using \NRSur{} and found that, while the kick magnitude was only weakly constrained, the recoil direction was informative.
For the GWTC-4 and GWTC-5 catalogs, recoil kicks have been provided for a subset of events as part of a post-processed data release~\cite{LIGOScientific:2025slb,LIGOScientific:2026wfs}.

In this paper, we present recoil-kick velocity estimates for all events reported up to the GWTC--5 catalog~\cite{LIGOScientific:2018mvr,LIGOScientific:2020ibl,LIGOScientific:2021usb,KAGRA:2021vkt,LIGOScientific:2025slb,LIGOScientific:2026ctl,LIGOScientific:2026wfs,LIGOScientific:2025brd,LIGOScientific:2025rid} using a consistent analysis framework. In addition, we compute recoil velocities for the 12 LVK intermediate-mass black hole (IMBH) candidates presented in Ref.~\cite{LIGOScientific:2021tfm} and subsequently analyzed in Ref.~\cite{Ruiz-Rocha:2025yno}. Our final sample consists of 286 GW events and candidate binaries, including 53 events from GWTC--2.1, 34 from GWTC--3, 84 from GWTC--4, 103 from GWTC--5, and 12 LVK IMBH candidates, making this the most comprehensive recoil-kick analysis of GW sources performed to date.
Beyond estimating recoil kicks for individual events, we also infer recoil-kick distributions from the population properties of BBH mergers reported in GWTC--3, GWTC--4, and GWTC--5. Using both event-level and population-level recoil distributions, we investigate the retention of merger remnants in different astrophysical environments, quantify the impact of recoil-induced displacements and dynamical-friction return times, and assess the conditions under which retained remnants can participate in hierarchical mergers. In particular, we show that remnant retention alone does not necessarily imply efficient hierarchical growth, as recoil-driven excursions from dense cluster cores can substantially suppress the probability of subsequent mergers.

We provide the details of the recoil model in Section~\ref{sec:model}. We present recoil-kick estimates for individual GW events in Section~\ref{sec:event_kick} and infer recoil-kick distributions from BBH population models in Section~\ref{sec:population_kick}. Finally, we discuss the astrophysical implications of these results in Section~\ref{sec:astro}. Our results are publicly available at \href{https://github.com/tousifislam/GWTCKick}{https://github.com/tousifislam/GWTCKick}. The software framework used to compute recoil kicks and perform the astrophysical analyses presented in this work is publicly available through the \textcolor{linkcolor}{\texttt{gwGenealogy}}\footnote{\href{https://github.com/tousifislam/gwGenealogy}{https://github.com/tousifislam/gwGenealogy}} package.

\section{Model for the recoil kick}
\label{sec:model}
A GW signal from a quasi-circular BBH merger is fully described by 15 parameters, consisting of eight intrinsic and seven extrinsic parameters~\cite{Maggiore:2007ulw,Maggiore:2018sht}. The intrinsic parameters include the component masses $m_1$ and $m_2$, the dimensionless spin magnitudes $|\chi_1|$ and $|\chi_2|$, the spin tilt angles $\theta_1$ and $\theta_2$ measured relative to the orbital angular momentum, and the azimuthal spin angles $\phi_1$ and $\phi_2$. The extrinsic parameters characterize the source location, orientation, and the reference time and phase of the binary. The recoil kick depends only on the intrinsic parameters of the binary. Moreover, it is insensitive to the absolute mass scale and instead depends on the mass ratio, defined as $q := m_2 / m_1$. The mapping between the intrinsic parameters and the recoil velocity can be written schematically as
\begin{equation}
    \{ q, |\chi_1|, |\chi_2|, \theta_1, \theta_2, \phi_1, \phi_2, X_{\rm ref} \} \mapsto v_{\rm kick},
\end{equation}
where $X_{\rm ref}$ denotes the reference point at which the spin vectors are specified, such as a reference orbital separation, reference frequency, or equivalently a reference time before merger.

In this work, we construct the mapping using a combination of \NRSur{} and \HLZ{}. The \NRSur{} model is formally trained for BBH mergers with mass ratios in the range $0.25 \leq q \leq 1$ and spin magnitudes $|\chi_{1,2}| \leq 0.8$. In practice, it is often extrapolated and applied within the extended domain $0.1667 \leq q \leq 1$ and $|\chi_{1,2}| \leq 1$, although extrapolation beyond this range is not recommended, as the model accuracy may degrade. For $q \leq 0.1667$, the binary lies entirely outside the commonly adopted extrapolation range of \NRSur{}. In this regime the surrogate predictions rely on extrapolation of the underlying Gaussian-process model beyond its validated domain and are therefore not expected to provide reliable recoil estimates.
By contrast, \HLZ{} is a semi-analytical model based on PN-inspired recoil scalings and calibrated to NR simulations. Unlike \NRSur{}, the model can be evaluated throughout the full parameter space of mass ratios and spins. While its accuracy outside the calibration region is not rigorously established, its analytic PN-motivated structure provides a well-behaved extrapolation to larger mass ratios and spin magnitudes.

We therefore adopt the following \textit{practical} prescription: we use \NRSur{} whenever $q \geq 0.1667$, and \HLZ{} otherwise. Consequently, for a given GW event whose inferred posterior contains samples on both sides of $q = 0.1667$, we evaluate the recoil velocity using \HLZ{} for samples with $q \leq 0.1667$ and \NRSur{} for samples with $q \geq 0.1667$. While this procedure introduces model-dependent systematic uncertainties associated with the transition between the two recoil prescriptions, comparisons between the default, HLZ-only, and NRSur-only analyses indicate that the resulting recoil distributions are generally robust to the choice of prescription (see Sec.~\ref{sec:event_kick}). Similar prescriptions have also been adopted in Refs.~\cite{Yang:2019cbr,Mahapatra:2024qsy}. Unless stated otherwise, this prescription is adopted as the default throughout the paper.
We access the \NRSur{} model through the \textcolor{linkcolor}{\texttt{surfinBH}}\footnote{\href{https://github.com/vijayvarma392/surfinBH/}{https://github.com/vijayvarma392/surfinBH/}} package, whereas the \HLZ{} model is accessed through the \textcolor{linkcolor}{\texttt{gwModels}}\footnote{\href{https://github.com/tousifislam/gwModels}{https://github.com/tousifislam/gwModels}} package.

Another important aspect is the choice of the appropriate reference frame $X_{\rm ref}$ for the initial binary parameters required by the recoil models. For \NRSur{}, this frame is defined at $t = -100M$ in geometric units (where $G = c = 1$ and $M$ denotes the total mass of the binary). 
In contrast, the \HLZ{} prescription requires the binary parameters to be specified close to merger. Following the conventions adopted in the \textcolor{linkcolor}{\texttt{precession}} package\footnote{\href{https://github.com/dgerosa/precession/}{https://github.com/dgerosa/precession/}}, we choose to specify the binary parameters at a separation of $R_{\rm sep}=10M$.

The posterior samples used as input to our framework typically provide binary parameters defined at a reference GW frequency $f_{\rm ref}^{\rm post} = 20\,\mathrm{Hz}$ for most events, although this value differs for a small subset of events. Therefore, for each posterior sample, we evolve the initial binary parameters, in particular the spin angles, from $f_{\rm ref}^{\rm post}$ to the reference frame appropriate for the chosen recoil model. 
For the \HLZ{} prescription, we evolve the binary parameters from
$f_{\rm ref}^{\rm post}$ to $R_{\rm sep}=10M$ using the orbit-averaged
PN evolution implemented in the \texttt{precession} package, which
includes spin-precession effects through 2PN order and radiation-reaction
corrections through 3.5PN (2PN) order for nonspinning (spinning) terms~\cite{Kidder:1995zr}:
\begin{equation}
f_{\rm ref}^{\rm post}
\xrightarrow{\rm PN\ spin\ evolution}
R_{\rm sep}=10M
\xrightarrow{\rm HLZ}
v_{\rm kick}.
\end{equation}
For \NRSur{}, we use the native \textcolor{linkcolor}{lalsimulation}\footnote{\href{https://lscsoft.docs.ligo.org/lalsuite/lalsuite/index.html}{https://lscsoft.docs.ligo.org/lalsuite/lalsuite/index.html}} spin evolution implemented in \textcolor{linkcolor}{\texttt{surfinBH}}. When the posterior reference frequency lies within the surrogate's training range, the surrogate-based spin evolution is applied directly. Otherwise, the binary parameters are first evolved from $f_{\rm ref}^{\rm post}$ to the beginning of the surrogate regime using \texttt{SpinTaylorT4} PN equations, after which the native surrogate spin evolution is employed to obtain the binary parameters at the reference time required by \NRSur{}:
\begin{equation}
f_{\rm ref}^{\rm post}
\xrightarrow{\rm SpinTaylorT4}
\text{surrogate regime}
\xrightarrow{\rm surrogate\ spin\ evolution}
v_{\rm kick}.
\end{equation}
This procedure ensures that we utilize the \NRSur{} model whenever possible.
For additional checks, we use \gwModelXP{} (for precessing binaries with isotropic spins) from \textcolor{linkcolor}{\texttt{gwModels}} package.

\begin{figure}
    \centering
    \includegraphics[width=\columnwidth]{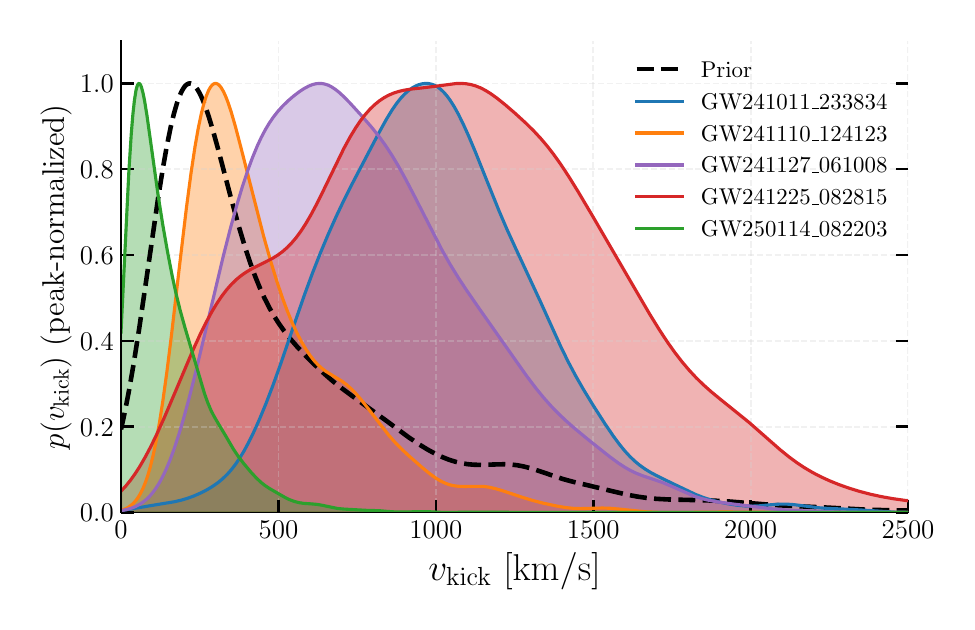}
    \caption{\textit{Recoil kicks for selected GWTC--5 events}. We show the inferred recoil kick velocity posteriors for a subset of recently reported GWTC--5 events: GW241011\_233834 (blue), GW241110\_124123 (orange), GW241127\_061008 (red), GW241225\_082815 (purple), and GW250114\_082203 (green). For reference, we also show the corresponding kick prior distributions (black dashed line). More details are in Section~\ref{sec:event_kick}.}
    \label{fig:O4bevents}
\end{figure}

\section{Recoil kick inference for GW events}
\label{sec:event_kick}
We first infer recoil kicks for individual GW events using publicly available posterior samples. The input to our framework are the inferred initial binary source parameters for each GW event. These inferences are obtained using a variety of gravitational waveform models. Across the GWTC catalogs, events are routinely analyzed with phenomenological waveform models (e.g., \texttt{IMRPhenomXPHM}~\cite{Pratten:2020ceb} for GWTC-2.1 and GWTC-3, and \texttt{IMRPhenomXPHM-SpinTaylor}~\cite{Colleoni:2024knd} for GWTC-4 and GWTC-5). For many events, parameter-estimation results obtained with effective-one-body (EOB) models are also available (e.g., \texttt{SEOBNRv4PHM}~\cite{Ossokine:2020kjp} for GWTC-2.1 and GWTC-3, and \texttt{SEOBNRv5PHM}~\cite{Ramos-Buades:2023ehm}) for GWTC-4 and GWTC-5). In addition, a smaller subset of events has been analyzed using NR surrogate waveform models such as \texttt{NRSur7dq4}.
Based on the relative accuracy of these waveform families, we adopt the following selection procedure. For each event, we first check whether posterior samples obtained with an NR surrogate model are available in Refs.~\cite{LIGOScientific:2025slb,Islam:2023zzj}. If so, we use the NR surrogate posteriors as input to our recoil framework. If not, we use posterior samples derived from EOB models when available. Otherwise, we default to posterior samples obtained with phenomenological waveform models. For the LVK IMBH candidates, we have utilized the inferred binary source parameters provided in Ref.~\cite{Ruiz-Rocha:2025yno} and choose to use the \texttt{SEOBNRv4PHM} posteriors as our input to the recoil model. These choices are based on the mismatch studies performed against NR simulations~\cite{Pratten:2020ceb,Yu:2023lml,MacUilliam:2024oif}.

To assess whether the inferred recoil kick posterior is informative, we compare it with the corresponding kick prior. Since the recoil velocity is a derived quantity, its prior is induced by the assumed priors on the underlying binary source parameters. Typically, uniform priors are adopted for the component masses (which determine the prior on the mass ratio $q$) and for the spin magnitudes $|\chi_{1,2}|$. The spin orientations are sampled from an isotropic distribution, with $\cos\theta_{1,2} \sim \mathcal{U}(-1,1)$ and $\phi_{1,2} \sim \mathcal{U}(0,2\pi)$~\cite{Romero-Shaw:2020owr,Veitch:2014wba}. We draw binary source parameters from these priors and propagate them through the recoil model to obtain the kick prior:
\begin{equation}
    \{ q^{\rm prior}, |\chi_1^{\rm prior}|, |\chi_2^{\rm prior}|, 
    \theta_1^{\rm prior}, \theta_2^{\rm prior}, 
    \phi_1^{\rm prior}, \phi_2^{\rm prior}, X_{\rm ref} \}
    \mapsto \{v_{\rm kick}^{\rm prior}\}.
\end{equation}
We further investigate whether the kick inference is driven primarily by the mass ratio and spin magnitudes or whether the spin angles also play a significant role. To this end, we recompute the recoil velocities by propagating the inferred values of $\{ q, |\chi_1|, |\chi_2| \}$ while replacing the inferred spin angles with angles drawn from the isotropic prior. We denote the resulting recoil as $v_{\rm kick}^{\rm iso}$:
\begin{equation}
    \{ q, |\chi_1|, |\chi_2|, 
    \theta_1^{\rm prior}, \theta_2^{\rm prior}, 
    \phi_1^{\rm prior}, \phi_2^{\rm prior}, X_{\rm ref} \}
    \mapsto \{v_{\rm kick}^{\rm iso}\}.
\end{equation}

\begin{figure}
    \centering
    \includegraphics[width=\columnwidth]{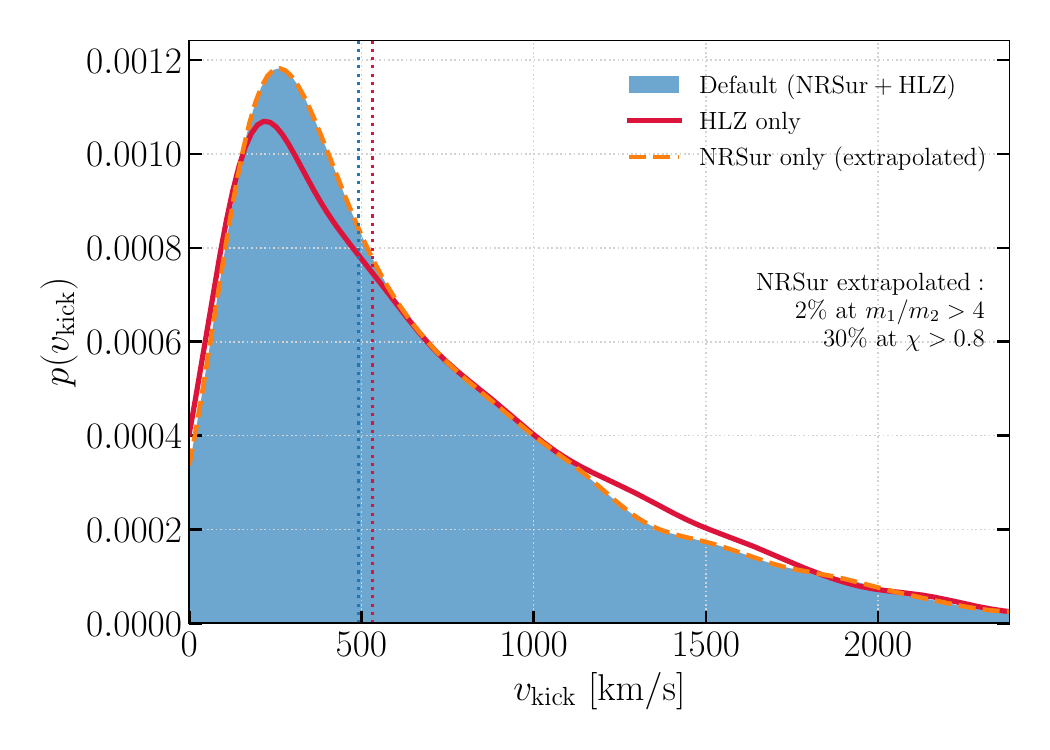}
    \caption{\textit{Robustness of recoil-kick inference to the choice of recoil prescription}. Using GW241229\_155844 as a representative example, we compare the recoil-kick distributions inferred using our default prescription (\NRSur{}+\HLZ{}; blue), the \HLZ{} prescription applied to all posterior samples (red), and the \NRSur{} prescription applied to all posterior samples, including extrapolation beyond its commonly adopted domain (orange). The inferred recoil distributions are broadly consistent across the different prescriptions, indicating that the principal conclusions of this work are not strongly sensitive to the choice of recoil model. More details are in Section~\ref{sec:event_kick}.}
    \label{fig:robustness_kick_methods}
\end{figure}

In Fig.~\ref{fig:errorbar_kicks}, we show the median and 90\% credible intervals of the recoil kick velocities inferred from publicly available posterior samples for all GWTC events, as well as LVK IMBH candidates. For visual clarity, the events are ordered by their median inferred recoil kick velocities. In Fig.~\ref{fig:select_gwtc4_kicks}, we show the recoil-kick velocity posteriors $v_{\rm kick}$ (our default estimate) for a subset of representative events whose inferred recoil posteriors exhibit noticeable deviations from the corresponding prior distributions. For comparison, we also show the kick posteriors $v_{\rm kick}^{\rm iso}$ obtained using our default prescription with isotropically distributed spin orientations, as well as results derived from \gwModel{}. The corresponding Jensen--Shannon divergence (JSD) values between the inferred recoil posterior and the prior, and between the default and isotropic-spin recoil posteriors, are indicated in Fig.~\ref{fig:select_gwtc4_kicks} for reference. We exclude events for which recoil-kick inferences have already been presented in Refs.~\cite{Islam:2023zzj,Mahapatra:2021hme}. 

Consistent with earlier findings~\cite{Varma:2020nbm,Islam:2023zzj,Mahapatra:2021hme}, we find that the kick velocity posteriors remain largely prior dominated for most events. Among the systems shown in Fig.~\ref{fig:select_gwtc4_kicks}, we obtain informative recoil constraints for GW231028\_153006, with $v_{\rm kick}=839^{+1018}_{-681}\,\mathrm{km\,s^{-1}}$, and GW231123\_135430, with $v_{\rm kick}=974^{+944}_{-760}\,\mathrm{km\,s^{-1}}$. For GW231114\_043211, we constrain the recoil velocity to $v_{\rm kick}=214^{+410}_{-98}\,\mathrm{km\,s^{-1}}$. For GW200210\_092254, we infer a moderately constrained recoil posterior of $v_{\rm kick}=88^{+88}_{-32}\,\mathrm{km\,s^{-1}}$. We also note that the nature of this event is currently ambiguous, as it may correspond either to a BBH merger or to a compact binary consisting of a black hole and a neutron star. 

We show the recoil-kick posteriors for a subset of GWTC--5 events in Fig.~\ref{fig:O4bevents}. The inferred recoil velocities are
\begin{align*}
\text{GW241011\_233834} : \quad & v_{\rm kick} = 974^{+555}_{-466}\,\mathrm{km\,s^{-1}}, \\
\text{GW241110\_124123} : \quad & v_{\rm kick} = 394^{+582}_{-207}\,\mathrm{km\,s^{-1}}, \\
\text{GW241127\_061008} : \quad & v_{\rm kick} = 732^{+706}_{-426}\,\mathrm{km\,s^{-1}}, \\
\text{GW241225\_082815} : \quad & v_{\rm kick} = 1053^{+868}_{-760}\,\mathrm{km\,s^{-1}}, \\
\text{GW250114\_082203} : \quad & v_{\rm kick} = 115^{+301}_{-95}\,\mathrm{km\,s^{-1}}.
\end{align*}
We find that, while the recoil posterior for GW241110\_124123 closely resembles the corresponding prior distribution, the posteriors for GW241011\_233834, GW241127\_061008, GW241225\_082815, and GW250114\_082203 exhibit noticeable deviations from their priors. Among these events, GW241225\_082815 exhibits the largest median inferred recoil velocity, followed by GW241011\_233834.

\begin{figure*}
    \centering
    \includegraphics[width=0.9\textwidth]{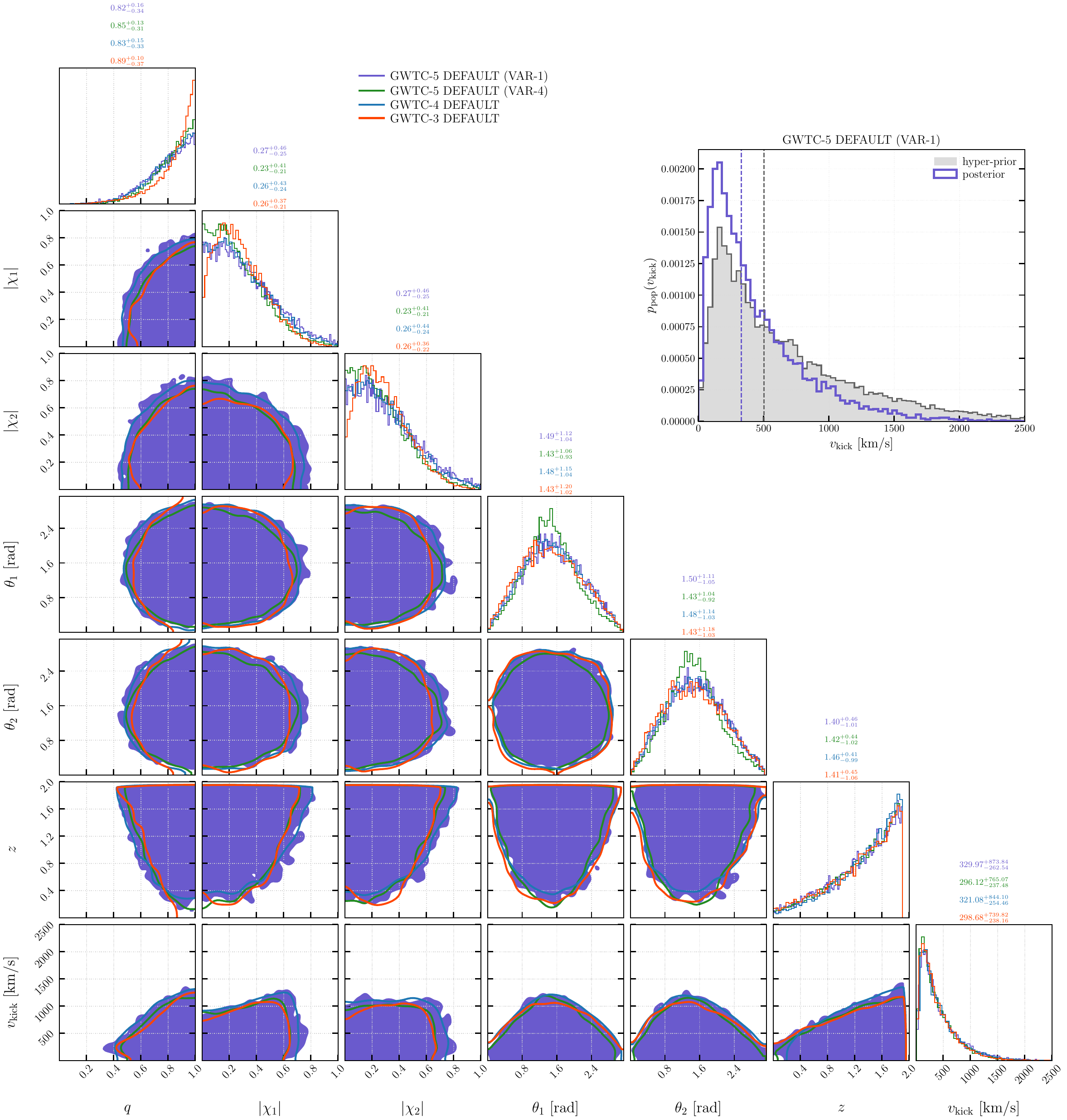}
    \caption{\textit{Population-level recoil kicks inferred from GWTC catalogs}. We show the inferred population distributions of the mass ratio $q$, spin magnitudes $|\chi_1|$ and $|\chi_2|$, spin tilt angles $\theta_1$ and $\theta_2$, redshift $z$, and recoil kick velocity $v_{\rm kick}$ obtained by propagating the LVK population posteriors through our recoil framework. Results are shown for the GWTC--3 (orange), GWTC--4 (blue), GWTC--5 VAR--4 (green), and GWTC--5 VAR--1 (violet) population models. The upper-right inset panel compares the recoil-kick distribution inferred from the GWTC--5 VAR--1 population posterior (violet) with the corresponding hyper-prior (gray). The inferred population recoil distributions are broadly consistent across catalogs, with median recoil velocities of approximately $300\text{--}330 \,\mathrm{km\,s^{-1}}$. More details are in Section~\ref{sec:population_kick}.}
    \label{fig:pop_corner_population_models}
\end{figure*}

One potential source of systematic uncertainty in our analysis is the choice of recoil prescription used outside the calibration domain of \NRSur{}. To assess the impact of this choice, we compare the recoil distributions obtained using our default prescription (\NRSur{}+\HLZ{}), the \HLZ{} prescription applied to all posterior samples, and the \NRSur{} prescription applied to all posterior samples, including extrapolation beyond its commonly adopted domain. The results are shown in Fig.~\ref{fig:robustness_kick_methods}.
As a representative example, we consider GW241229\_155844. This event contains 15,770 posterior samples, of which only 2.3\% satisfy $m_1/m_2 > 4$, while 30.0\% have at least one spin magnitude exceeding $0.8$. We find that the inferred recoil distributions are very similar across the different prescriptions. Using our default prescription, we obtain $v_{\rm kick}=491^{+1246}_{-374}\,\mathrm{km\,s^{-1}}$, while the \HLZ{}-only analysis yields $v_{\rm kick}=532^{+1256}_{-426}\,\mathrm{km\,s^{-1}}$. The corresponding \NRSur{}-only analysis is nearly indistinguishable from the default result. Similar behavior is observed for the majority of events considered in this work. 
Furthermore, most GWTC events have negligible posterior support for mass ratios below $q=0.1667$. For example, among the 116 BBH mergers in GWTC--3 and GWTC--4 considered in this work, 57.8\% have no posterior support at $q<0.1667$. For the remaining 42.2\% of events, the median fraction of posterior samples satisfying $q<0.1667$ is only 0.29\%. Consequently, only a very small fraction of the posterior samples require the \HLZ{} prescription, and the inferred recoil distributions are generally dominated by the \NRSur{} model. We therefore conclude that the recoil distributions inferred in this work are not strongly sensitive to the specific recoil prescription adopted.

\section{Recoil kick inference from population properties}
\label{sec:population_kick}
The event-level recoil posteriors discussed in Section~\ref{sec:event_kick} provide constraints on individual mergers. We now turn to population-level recoil kicks inferred from the population properties of BBH mergers. In particular, we use the population posteriors released by the LVK Collaboration for GWTC--3~\cite{KAGRA:2021duu}, GWTC--4~\cite{LIGOScientific:2025pvj}, and GWTC--5~\cite{LIGOScientific:2026ctl}, together with the corresponding hyper-priors, and propagate the inferred population distributions through our recoil framework. For each hyper-posterior sample $\Lambda_i$, we draw BBH source parameters
\[
\Theta=(q,|\chi_1|,|\chi_2|,\theta_1,\theta_2)
\]
from the corresponding population model $p(\Theta|\Lambda_i)$. The LVK population models do not explicitly model the azimuthal spin angles $\phi_1$ and $\phi_2$; therefore, for each draw we sample these angles independently from the isotropic prior, $\phi_1,\phi_2\sim U(0,2\pi)$. The resulting spin configuration is then propagated through the recoil prescription described in Section~\ref{sec:model} to obtain recoil velocities $v_{\rm kick}$. Repeating this procedure over the full set of hyper-posterior samples yields Monte-Carlo samples from the population recoil distribution,
\begin{equation}
\begin{split}
p_{\rm pop}(v_{\rm kick})
&=
\int p(v_{\rm kick}|\Theta,\phi_1,\phi_2)\,
p(\phi_1)\,
p(\phi_2) \\
&\quad \times
p(\Theta|\Lambda)\,
p(\Lambda|d)\,
d\phi_1\,d\phi_2\,d\Theta\,d\Lambda .
\end{split}
\end{equation}
thereby marginalizing over uncertainties in the population hyper-parameters. The resulting recoil distribution can be directly compared against the distribution implied by the corresponding hyper-prior.

The inferred population properties of BBH mergers can depend on the choice of population model. Moreover, the default population models adopted by the LVK Collaboration have evolved between GWTC--3, GWTC--4, and GWTC--5. To facilitate a meaningful comparison across catalogs, we restrict our analysis to the default population models that treat the mass-ratio distribution, spin-magnitude distribution, spin-orientation distribution, and redshift distribution as independent components. Specifically, we use the GWTC--3 default model (Power-Law + Peak masses, Beta spin magnitudes, Gaussian+isotropic spin tilts, and a Power-Law redshift distribution), the GWTC--4 default model (Broken-Power-Law Two-Peaks masses, Gaussian spin magnitudes, Gaussian+isotropic spin tilts, and a Power-Law redshift distribution), and the GWTC--5 default models VAR--1 and VAR--4. Both GWTC--5 models employ the TwoPeakBrokenPowerLaw mass model, Gaussian spin magnitudes, Gaussian+isotropic spin tilts, and a Power-Law redshift distribution, but differ in the specific parameterization adopted by the LVK Collaboration~\cite{LIGOScientific:2026ctl}. For both the spin-magnitude and spin-tilt distributions, these population models allow the primary and secondary black holes to be described by the same functional form but with independent sets of hyper-parameters.

We show the inferred population distributions of the mass ratio $q$, spin magnitudes $|\chi_1|$ and $|\chi_2|$, spin tilt angles $\theta_1$ and $\theta_2$, redshift $z$, and recoil kick velocity $v_{\rm kick}$ for the GWTC--3, GWTC--4, and GWTC--5 population models in Fig.~\ref{fig:pop_corner_population_models}. Despite differences in the underlying population prescriptions, the inferred population properties remain broadly consistent across catalogs. In particular, the inferred recoil-kick distributions exhibit only modest variations, with median recoil velocities ranging from approximately $300$ to $330~{\rm km\,s^{-1}}$. We further compare the recoil-kick distribution inferred from the GWTC--5 VAR--1 population posterior with the corresponding hyper-prior. We find that the posterior distribution favors systematically lower recoil velocities than expected from the hyper-prior, with the median recoil velocity decreasing from approximately $500~{\rm km\,s^{-1}}$ for the hyper-prior to approximately $330~{\rm km\,s^{-1}}$ for the posterior population. In any case, these velocities substantially exceed the escape speeds of typical globular clusters and dwarf galaxies.

\begin{figure}
    \includegraphics[width=\columnwidth]{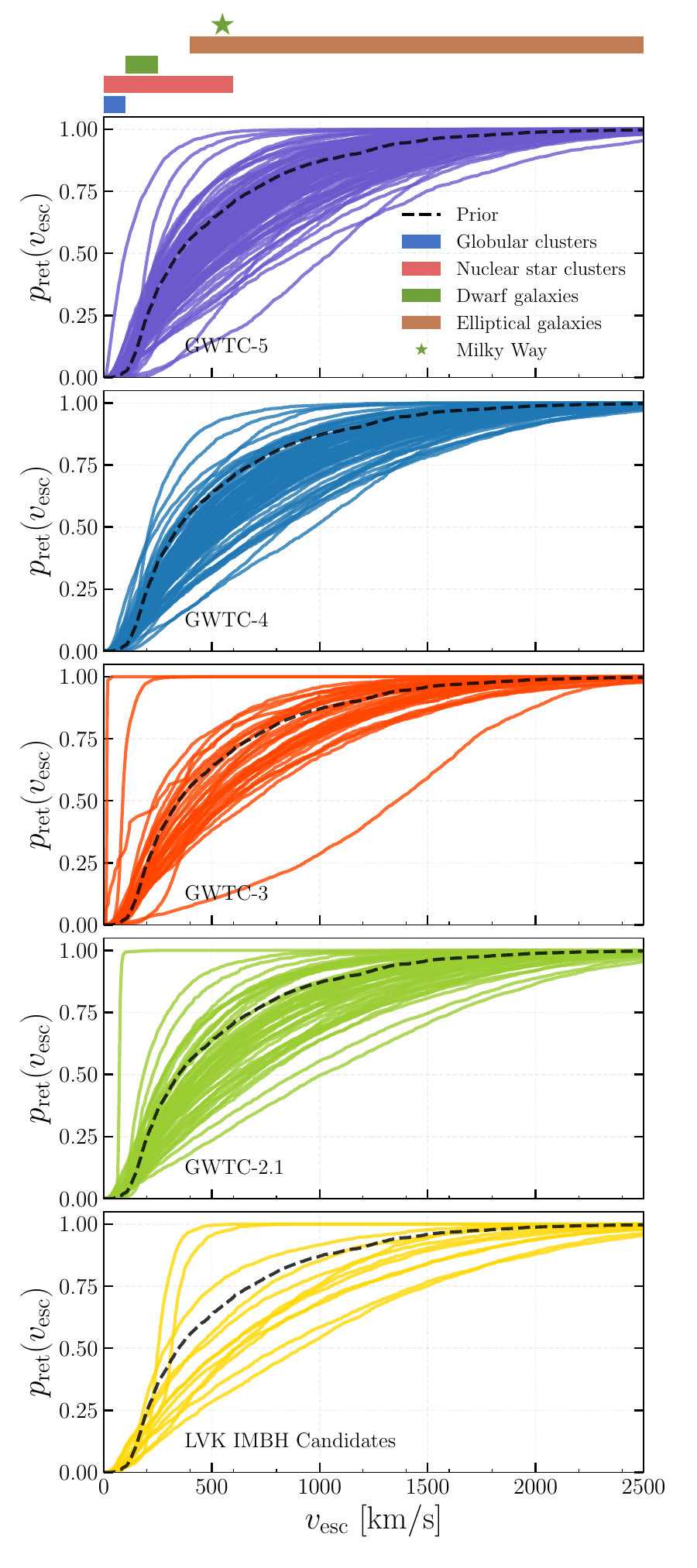}
    \caption{\textit{Retention probability of individual GW events}. We show the retention probability of all BBH merger remnants for events in GWTC--5 (violet), GWTC--4 (blue), GWTC--3 (orange-red), GWTC--2.1 (green), and LVK candidate IMBH mergers (yellow) as a function of the escape velocity in the range $[0, 2500]\,\mathrm{km\,s^{-1}}$. For reference, we also indicate representative escape velocity ranges for globular clusters, nuclear star clusters, dwarf galaxies, elliptical galaxies, and the Milky Way on top. Additionally, we show the induced prior for the retention probability as a function of the escape velocities as black dashed lines. More details are in Section~\ref{sec:astro_events_pret}.}
    \label{fig:gwtc_retention_probability}
\end{figure}

\section{Astrophysical implications}
\label{sec:astro}
We now discuss the astrophysical implications of our recoil-kick estimates obtained from both individual GW events and the inferred population properties. We choose to consider both because they provide complementary information. The recoil-kick posteriors inferred from individual GW events are directly tied to specific mergers and are largely insensitive to assumptions about the underlying BBH population. However, they do not provide a complete picture of the astrophysical population because selection effects are not accounted for. In contrast, recoil-kick distributions inferred from population models naturally incorporate the effects of selection bias and therefore provide a more representative description of the underlying BBH population. These population-level results, however, can depend on the adopted population model. We therefore consider both event-level and population-level recoil estimates in the following analysis, as together they provide a more robust picture of the astrophysical consequences of BBH recoil kicks.

\subsection{Astrophysical implications of individual GW-event kicks}
\label{sec:astro_events}

\subsubsection{Retention probability of GWTC remnant black holes}
\label{sec:astro_events_pret}
We begin by computing the retention probability of the remnant black holes associated with the observed GW events as a function of the escape velocity $v_{\rm esc}$ in the range $0$--$2500\,\mathrm{km\,s^{-1}}$ (Fig.~\ref{fig:gwtc_retention_probability}).
Given the posterior probability distribution $p(v_{\rm kick})$ of the recoil velocity for a remnant black hole, we compute the retention probability $p_{\rm ret}(v_{\rm esc})$ as a function of the escape velocity $v_{\rm esc}$ as the cumulative distribution function (CDF):
\begin{equation}
p_{\rm ret}(v_{\rm esc}) 
= \int_{0}^{v_{\rm esc}} p(v_{\rm kick}) \, \mathrm{d}v_{\rm kick}.
\label{eq:retention_cdf}
\end{equation}
To place the agnostic escape velocity values in an astrophysical context, we adopt representative escape velocity ranges for different host environments: globular clusters ($0$--$150\,\mathrm{km\,s^{-1}}$), nuclear star clusters (NSCs) ($0$--$1100\,\mathrm{km\,s^{-1}}$), dwarf galaxies ($100$--$250\,\mathrm{km\,s^{-1}}$), and elliptical galaxies ($400$--$2100\,\mathrm{km\,s^{-1}}$)~\cite{Merritt:2004xa,Antonini:2016gqe}. For additional context, we also indicate the escape velocity of the Milky Way ($\sim500\,\mathrm{km\,s^{-1}}$)~\cite{2018A&A...616L...9M}. These reference values allow us to interpret the retention probabilities of individual merger remnants within realistic astrophysical settings.
We find that, aside from a small number of outliers, the retention
probability distributions are broadly similar across catalogs.

\begin{figure}
    \includegraphics[width=\columnwidth]{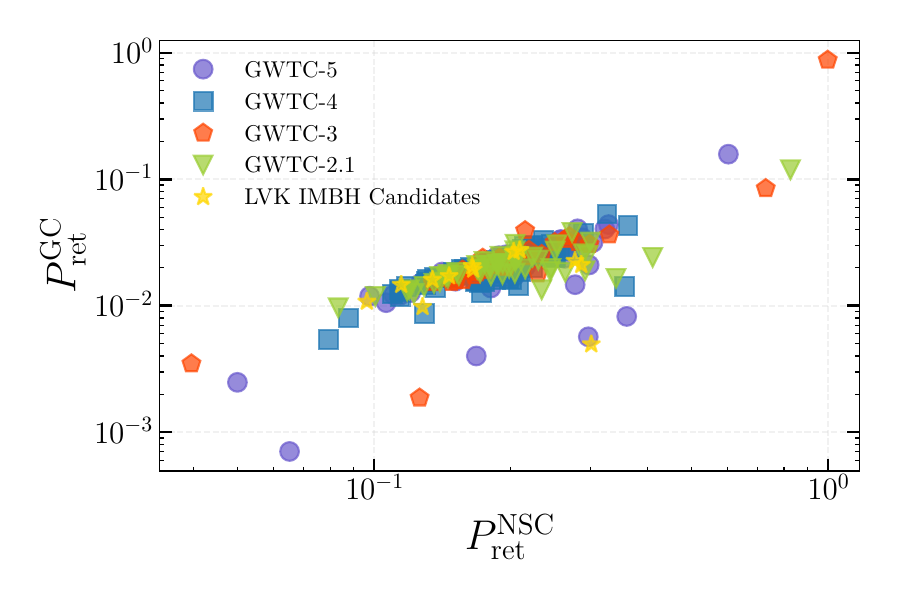}
    \includegraphics[width=\columnwidth]{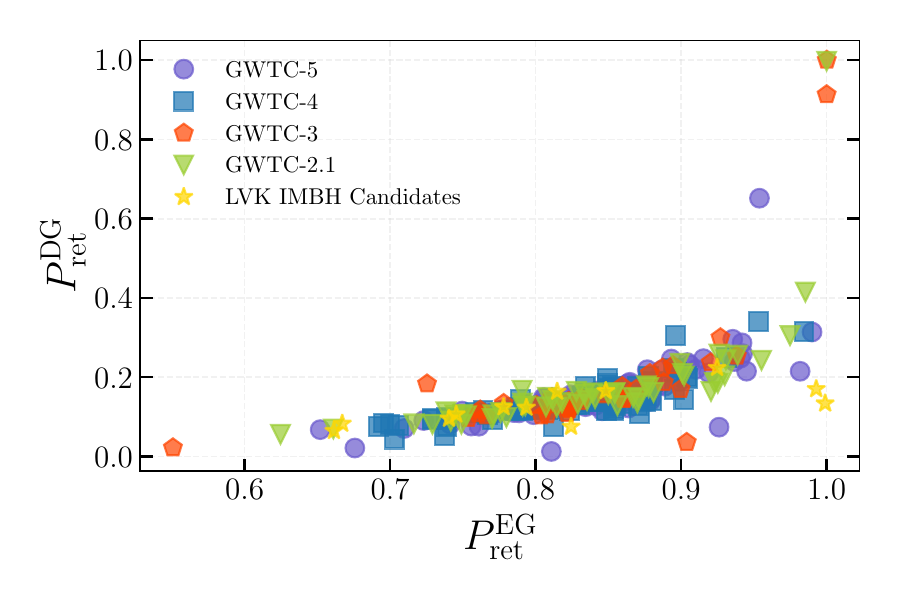}
    \caption{\textit{Overall retention probability of all GW events in different host environment}. \textbf{Upper panel: }We show the retention probability of BBH merger remnants for events in GWTC--5 (violet), GWTC--4 (blue), GWTC--3 (orange-red), GWTC--2.1 (green), and LVK candidate IMBH mergers (yellow), assuming the mergers occurred in globular clusters (GC) and nuclear star clusters (NSC). \textbf{Lower panel: } Same but for dwarf galaxies (DG) and elliptical galaxies (EG). More details are in Section~\ref{sec:astro_events_pret}.}
    \label{fig:gc_nsc_dg_ng_retention_probability}
\end{figure}

To obtain an \emph{overall} retention probability for a given host population, we marginalize over the distribution of escape speeds in that environment:
\begin{equation}
P_{\rm ret} =
\int_{0}^{\infty} p(v_{\rm kick}) 
\left[ \int_{v_{\rm kick}}^{\infty} p(v_{\rm esc}) \, \mathrm{d}v_{\rm esc} \right]
\mathrm{d}v_{\rm kick}.
\label{eq:pret_marginalized}
\end{equation}
In practice, we evaluate Eq.~(\ref{eq:pret_marginalized}) numerically by constructing a kernel density estimate of $p(v_{\rm kick})$ from posterior samples and computing the survival probability
$P(v_{\rm esc} > v_{\rm kick}) = 1 - F(v_{\rm kick})$.
For globular clusters and nuclear star clusters, we model the escape speeds using a log-normal distribution following Ref.~\cite{Antonini:2016gqe}. Specifically, we assume that $\log_{10}(v_{\rm esc}/\mathrm{km\,s^{-1}})$ is normally distributed with mean $\mu_{\log_{10}}$ and standard deviation $\sigma_{\log_{10}}$, which implies
\begin{equation}
F_{\rm esc}(v_{\rm esc})
= \Phi\!\left(\frac{\ln v_{\rm esc} - \mu_{\ln}}{\sigma_{\ln}}\right),
\label{eq:lognormal_vesc_cdf}
\end{equation}
where $\Phi$ is the standard normal CDF with $\mu_{\ln} = \mu_{\log_{10}}\ln 10$ and $\sigma_{\ln} = \sigma_{\log_{10}}\ln 10$. We use $(\mu_{\log_{10}},\sigma_{\log_{10}})=(2.2,0.36)$ for NSCs and $(1.5,0.30)$ for GCs. For dwarf galaxies (DG) and elliptical galaxies (EG), we assume a uniform distribution of escape velocities within observationally motivated bounds mentioned before:
\begin{equation}
p(v_{\rm esc}) =
\begin{cases}
\dfrac{1}{v_{\max} - v_{\min}}, & v_{\min} \le v_{\rm esc} \le v_{\max}, \\
0, & \text{otherwise}.
\end{cases}
\label{eq:uniform_vesc}
\end{equation}
In Fig.~\ref{fig:gc_nsc_dg_ng_retention_probability}, we show the resulting retention probabilities for all events, color-coded by catalog. For globular clusters, we find that most retention probabilities lie in the range $[0.01,0.08]$, whereas for nuclear star clusters they typically span $[0.1,0.4]$. Retained black holes may subsequently form binaries with other black holes in the host cluster and participate in hierarchical merger processes. We perform a similar calculation for dwarf and elliptical galaxies. The typical retention probabilities are found to lie in the range $0.05$--$0.25$ for dwarf galaxies and $0.7$--$0.97$ for elliptical galaxies (Fig.~\ref{fig:gc_nsc_dg_ng_retention_probability}). This trend is expected, as these environments are characterized by larger escape velocities. Our estimated retention probabilities broadly agree with the values reported in Refs.~\cite{Mahapatra:2021hme,Doctor:2021qfn}.

While we have mostly focused on different types of stellar clusters so far, another potential host environment for BBH mergers is AGNs~\cite{Yang:2019cbr}. Unlike NSCs, which are stellar-dynamical systems, AGN disks are gas-dominated environments surrounding accreting supermassive black holes. The much larger escape velocities expected in AGNs ($\sim 1000$--$3000\,\mathrm{km\,s^{-1}}$) imply that recoil kicks are unlikely to eject the remnant black holes. Consequently, if any of the GWTC mergers occur in AGNs, the merger remnants are expected to be retained and may participate in hierarchical mergers.

\subsubsection{Wandering GWTC remnant black holes}
\label{sec:wanderer}
Globular clusters have relatively low escape velocities ($0$--$150\,\mathrm{km\,s^{-1}}$), whereas their host galaxies typically have much larger escape velocities ($\sim500$--$700\,\mathrm{km\,s^{-1}}$) for Milky Way--like systems~\cite{Merritt:2004xa,Antonini:2016gqe,2018A&A...616L...9M}. On the other hand, as discussed in Section~\ref{sec:astro_events_pret}, many of the inferred recoil kick velocities exceed $250\,\mathrm{km\,s^{-1}}$. Based on the retention probabilities presented in Section~\ref{sec:astro_events_pret}, we infer that the probability for a GWTC merger remnant to be ejected from its host globular cluster, if the merger occurred there, is $\sim0.9$.
Most of these merger remnants are therefore expected to be ejected from their parent clusters while remaining gravitationally bound to their host galaxies. As a result, these black holes would wander through the galactic halo. Such isolated remnant black holes may produce microlensing events~\cite{2023ApJ...955..116L}. In rare cases, two wandering black holes could encounter each other and form a binary through dynamical capture; however, the probability of such events is expected to be extremely small because of the low densities in galactic halos.
In the case of dwarf galaxies, the situation can be different because their escape velocities are typically much smaller ($100$--$250\,\mathrm{km\,s^{-1}}$)~\cite{Merritt:2004xa}. In such environments, a significant fraction ($0.75$--$0.95$) of GWTC merger remnants may exceed the galactic escape velocity and therefore be ejected entirely from their host galaxies, becoming intergalactic black holes.

\begin{figure}
    \centering
    \includegraphics[width=\columnwidth]{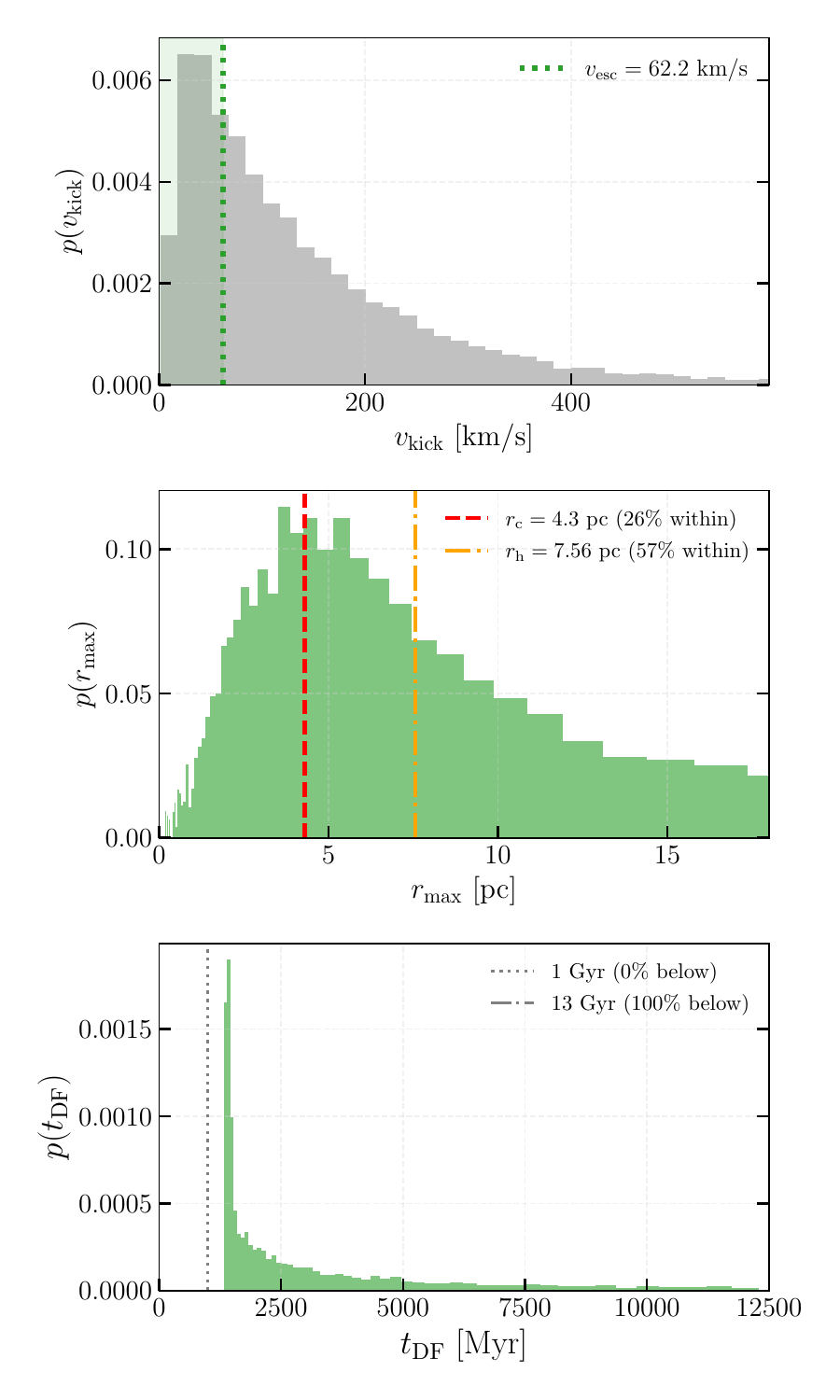}
    \caption{\textit{Example recoil displacement and dynamical-friction return time for GW250114\_082203 in an $\omega$-Centauri-like globular cluster}. The upper panel shows the recoil-kick posterior together with the cluster escape speed, $v_{\rm esc}=62.2\,\mathrm{km\,s^{-1}}$. The middle panel shows the distribution of maximum recoil displacements $r_{\max}$ for retained remnants, with the core radius $r_c=4.3\,\mathrm{pc}$ and half-mass radius $r_h=7.56\,\mathrm{pc}$ indicated by vertical lines. The lower panel shows the corresponding dynamical-friction return-time distribution $t_{\rm DF}$, with reference times of $1\,\mathrm{Gyr}$ and $13\,\mathrm{Gyr}$ shown for comparison. More details are in Section~\ref{sec:fallback}.}
    \label{fig:gw250114_omegacen_chandrasekhar}
\end{figure}

\subsubsection{Spatially displaced GWTC remnant black holes in globular clusters}
\label{sec:fallback}
Not all retained remnant black holes participate in hierarchical mergers with equal likelihood. The retention probabilities discussed in Section~\ref{sec:astro_events_pret} quantify whether a remnant remains gravitationally bound to its host environment, but retention alone does not guarantee subsequent dynamical interactions. Even when a remnant is retained, the recoil kick displaces it from its original location~\cite{2008ApJ...678..780G,Komossa:2008as}. To quantify this effect, we estimate both the maximum recoil displacement and the dynamical-friction return time of retained merger remnants~\cite{BinneyTremaine2008,Chandrasekhar1943}.

We model each globular cluster as a Plummer sphere with total mass $M_{\rm cl}$ and half-mass radius $r_h$~\cite{Plummer1911,BinneyTremaine2008}. The corresponding scale radius is $a = \frac{r_h}{1.305}$ and the gravitational potential is $\Phi(r)=-\frac{G M_{\rm cl}}{\sqrt{r^2+a^2}}$. The central escape speed is $v_{\rm esc}=\sqrt{\frac{2GM_{\rm cl}}{a}}$, while the speed required for a remnant to reach radius $r$ is \begin{equation} v(r)=\sqrt{\frac{2GM_{\rm cl}}{a}\left[1- \left(1+\frac{r^2}{a^2}\right)^{-1/2}\right]}. \end{equation} Defining $u \equiv \frac{v_{\rm kick}}{v_{\rm esc}}$, the maximum displacement of a retained remnant follows from energy conservation: \begin{equation} r_{\max}=a\,\sqrt{\left(1-u^2\right)^{-2}-1}. \label{eq:rmax} \end{equation} However, real globular clusters are tidally truncated. We therefore impose a tidal radius $r_t = 5\,r_h$, and count a remnant as cluster-retained only if it remains gravitationally bound and reaches an apocentre inside the tidal radius, i.e. if $v_{\rm kick}<v_{\rm esc}$ and $r_{\max}\le r_t$. To estimate the return time to the cluster core, we use Chandrasekhar dynamical friction~\cite{BinneyTremaine2008,Chandrasekhar1943}. The corresponding local dynamical-friction timescale is \begin{equation} t_{\rm DF,local}(r,v)=\frac{v^3}{4\pi G^2M_{\rm BH}\rho(r)\ln\Lambda\mathcal F(X)}, \end{equation} where $M_{\rm BH}$ is the remnant mass, $\mathcal F(X)=\operatorname{erf}(X)-\frac{2X}{\sqrt{\pi}}e^{-X^2}$ and $X=\frac{v}{\sqrt{2}\,\sigma(r)}$. We adopt the Coulomb logarithm $\ln\Lambda=2.5$. Here, $\rho(r)$ is the stellar mass density and $\sigma(r)$ is the one-dimensional velocity dispersion of the stellar background.

For the Plummer model, the density profile and one-dimensional velocity dispersion are
\begin{equation}
\rho(r)=\frac{3M_{\rm cl}}{4\pi a^3}\left(1+\frac{r^2}{a^2}\right)^{-5/2},
\end{equation}
and
\begin{equation}
\sigma^2(r)=\frac{GM_{\rm cl}}{6a}\left(1+\frac{r^2}{a^2}\right)^{-1/2}.
\end{equation}
Assuming a radial orbit, the velocity of the remnant at radius $r$ is
\begin{equation}
v_r(r)=v_{\rm esc}\sqrt{u^2-1+\left(1+\frac{r^2}{a^2}\right)^{-1/2}}.
\end{equation}
We define the return time as the orbit-averaged inverse local friction timescale,
\begin{equation}
t_{\rm DF}
=
\left\langle t_{\rm DF,local}^{-1}\right\rangle_{\rm orbit}^{-1}
=
\frac{\displaystyle\int_0^{r_{\max}}\frac{dr}{v_r(r)}}
{\displaystyle\int_0^{r_{\max}}\frac{dr}{v_r(r)\,t_{\rm DF,local}\!\left(r,v_r(r)\right)}}.
\label{eq:tdf}
\end{equation}
In practice, Eq.~(\ref{eq:tdf}) is evaluated numerically for each retained remnant. Smaller values of $r_{\max}$ and $t_{\rm DF}$ increase the likelihood that a retained remnant returns to the dense central regions of the cluster, where dynamical interactions and hierarchical mergers are expected to be most efficient.

\begin{figure}
    \includegraphics[width=\columnwidth]{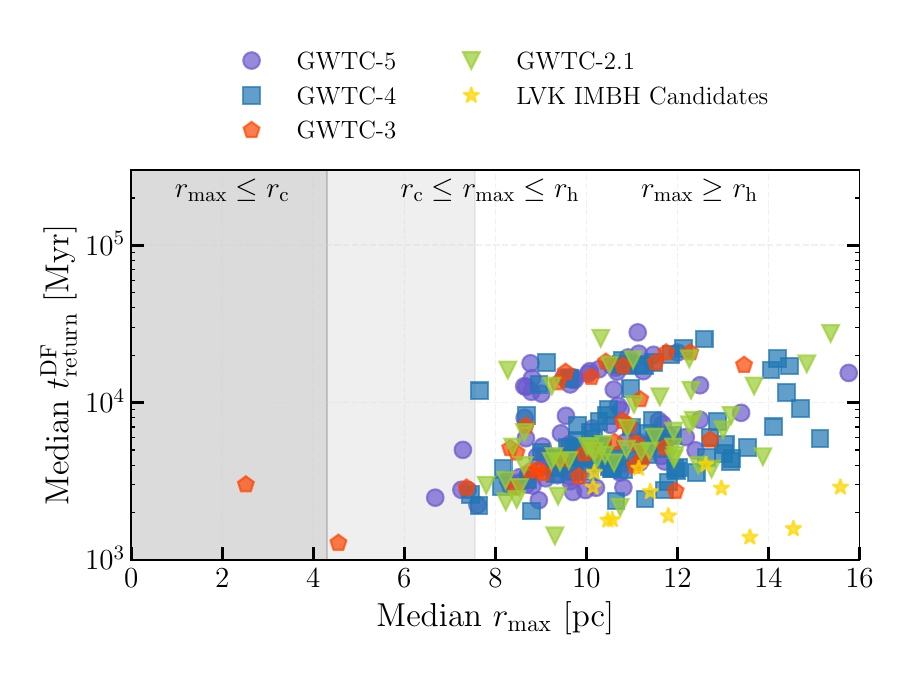}
    \includegraphics[width=\columnwidth]{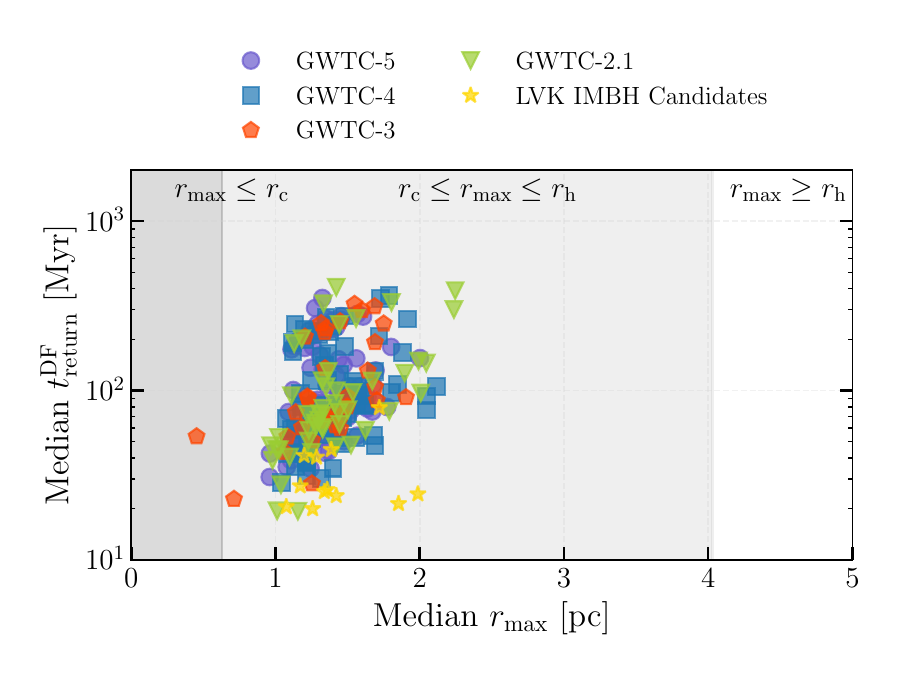}
    \caption{\textit{Understanding the effect of recoil kick in remnant displacement}. \textbf{Upper panel: }We show the median dynamical-friction return time $t_{\rm return}^{\rm DF}$ (in Myr) and the maximum displacement due to the recoil kick, $r_{\rm max}$, for BBH merger remnants corresponding to events in GWTC--5 (purple), GWTC--4 (blue), GWTC--3 (orange-red), GWTC--2.1 (green), and LVK candidate IMBH mergers (yellow), assuming the mergers occurred in globular clusters similar to $\omega$~Centauri. For reference, we indicate the regions $r_{\rm max} \leq r_c$ and $r_c \leq r_{\rm max} \leq r_h$ as shaded gray bands with different intensities. \textbf{Lower panel: }Same but for globular clusters similar to 47~Tuc. More details are in Section~\ref{sec:fallback}.}
    \label{fig:rmax_vs_treturn_omegacen}
\end{figure}

We demonstrate this framework using two well-studied Milky Way globular clusters, $\omega$~Centauri and 47~Tuc. Some more details are provided in Appendix~\ref{app:displacement}. Following Ref.~\cite{2018MNRAS.478.1520B}, we adopt $r_h=7.56\,\mathrm{pc}$ and $v_{\rm esc}=62.2\,\mathrm{km\,s^{-1}}$ for $\omega$~Centauri, and $r_h=4.03\,\mathrm{pc}$ and $v_{\rm esc}=47.4\,\mathrm{km\,s^{-1}}$ for 47~Tuc. For completeness, we also use the corresponding core radii $r_c=4.30\,\mathrm{pc}$ and $r_c=0.63\,\mathrm{pc}$ when interpreting the recoil displacements.
We select $\omega$~Centauri and 47~Tuc as representative examples spanning two extremes of Milky Way globular-cluster environments. $\omega$~Centauri is unusually massive and spatially extended, leading to large recoil excursions and long return times. In contrast, 47~Tuc is a compact, high-density cluster where retained remnants are expected to re-center more efficiently.

We first illustrate the calculation using GW250114\_082203 and an $\omega$~Centauri-like cluster. The results are shown in Fig.~\ref{fig:gw250114_omegacen_chandrasekhar}. The inferred recoil posterior is compared with the cluster escape speed, $v_{\rm esc}=62.2\,\mathrm{km\,s^{-1}}$. For retained posterior samples, we compute the corresponding apocentre $r_{\max}$ and dynamical-friction return time $t_{\rm DF}$. We find that $26\%$ of retained remnants remain within the core radius $r_c=4.3\,\mathrm{pc}$, while $57\%$ remain within the half-mass radius $r_h=7.56\,\mathrm{pc}$. The inferred return times are typically longer than $1\,\mathrm{Gyr}$ but shorter than a Hubble time, indicating that retained remnants can remain significantly displaced from the cluster centre for extended periods.

We then extend the calculation to all GW events. For each event, we first identify the subset of posterior samples that satisfy the retention criterion described above. For these retained samples, we compute the corresponding apocentre $r_{\max}$ and dynamical-friction return time $t_{\rm DF}$ assuming a globular cluster with properties similar to $\omega$~Centauri. The resulting distributions are shown in Fig.~\ref{fig:rmax_vs_treturn_omegacen} (upper panel).
We compare the inferred recoil displacement $r_{\max}$ with the cluster core radius $r_c$ and half-mass radius $r_h$. When $r_{\max}\le r_c$, the remnant remains confined to the dense core, where it is most likely to undergo subsequent dynamical interactions and participate in hierarchical mergers. Displacements in the range $r_c<r_{\max}\le r_h$ imply that retained remnants spend a substantial fraction of their orbital evolution outside the core, reducing the probability of interactions with other black holes. In cases where $r_{\max}>r_h$, the remnant undergoes a large-scale excursion through the cluster halo, making a prompt return to the core unlikely.
For $\omega$~Centauri, we find that the spatial-offset distribution of retained remnants is
\begin{align*}
r_{\rm max} \leq r_c \; (4.3\,\mathrm{pc}) &: \; 1 0.4\%, \\
r_c < r_{\rm max} \leq r_h \; (4.3\text{--}7.56\,\mathrm{pc}) &: \; 2.5\%, \\
r_{\rm max} > r_h \; (7.56\,\mathrm{pc}) &: \; 97.1\%.
\end{align*}
For remnants with $r_{\rm max} > r_h$, we find a median dynamical-friction return time of $5.22\times10^3\,\mathrm{Myr}$, with a 90\% credible interval of $[2.43\times10^3,\,1.98\times10^4]\,\mathrm{Myr}$.
These timescales are comparable to, or exceed, typical globular-cluster lifetimes, indicating that many retained remnants may remain displaced from the cluster centre for cosmologically significant periods of time.

We repeat the same analysis for 47~Tuc (Fig.~\ref{fig:rmax_vs_treturn_omegacen}, lower panel). In this case, the inferred recoil displacements are substantially smaller:
\begin{align*}
r_{\rm max} \leq r_c \; (0.63\,\mathrm{pc}) &: \; 0.4\%, \\
r_c < r_{\rm max} \leq r_h \; (0.63\text{--}4.03\,\mathrm{pc}) &: \; 99.6\%, \\
r_{\rm max} > r_h \; (4.03\,\mathrm{pc}) &: \; 0.0\%.
\end{align*}
Thus, while most retained remnants are displaced beyond the cluster core, essentially all remain confined within the half-mass radius. The corresponding dynamical-friction return times are substantially shorter than those obtained for $\omega$~Centauri, implying more efficient re-centering of retained remnants.

These results suggest that globular-cluster structure plays an important role in determining the likelihood of hierarchical mergers. In extended clusters such as $\omega$~Centauri, retained remnants are expected to spend long periods of time far from the cluster centre, suppressing subsequent dynamical interactions. In contrast, compact clusters such as 47~Tuc are significantly more effective at re-centering retained remnants, thereby enhancing the probability of hierarchical mergers.

\begin{figure}
    \includegraphics[width=\columnwidth]{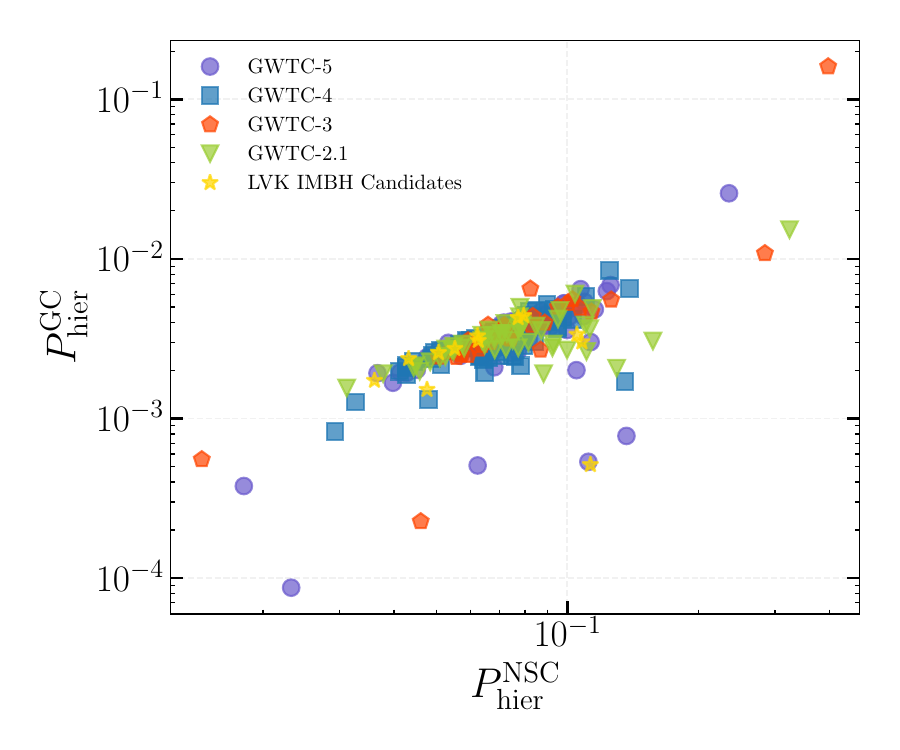}
    \caption{\textit{Probability of remnant participation in hierarchical mergers}. 
    We show the overall probability that BBH merger remnants participate in at least one hierarchical merger for events in GWTC--5 (purple), GWTC--4 (blue), GWTC--3 (orange-red), GWTC--2.1 (green), and LVK candidate IMBH mergers (yellow), assuming the mergers occur in globular clusters (GC) and nuclear star clusters (NSC). More details are in Section~\ref{sec:hierarchical}.}
    \label{fig:gc_nsc_hierarchical_probability}
\end{figure}

\subsubsection{Probability of GWTC remnant participation in hierarchical mergers}
\label{sec:hierarchical}
We now combine the information presented above to compute a crude estimate of the probability that a merger remnant from the GWTC events participates in at least one hierarchical merger in its host environment. We denote it as $P_{\rm hier}$. In particular, we focus on GCs and NSCs. In dense stellar systems, hierarchical mergers generally require three conditions. First, the remnant black hole must be retained within the cluster following the merger. Second, the retained remnant must form a new binary through subsequent dynamical interactions. Finally, the resulting binary must merge within the lifetime of the host environment. In addition, if the retained remnant receives a recoil kick that displaces it far from the cluster center, it may spend a significant fraction of its orbital evolution outside the dense core where dynamical interactions are most efficient. As a result, the probability of forming a new binary decreases as the dynamical-friction return time increases.
We can write
\begin{equation}
P_{\rm hier} = \int_0^\infty p(v_{\rm kick}) \, P(v_{\rm esc} > v_{\rm kick})
\, P_{\rm repeat} \, dv_{\rm kick},
\end{equation}
where $p(v_{\rm kick})$ is the posterior distribution of the recoil velocity inferred for the event, $P(v_{\rm esc} > v_{\rm kick})$ is the probability that the recoil velocity is smaller than the escape velocity of the host environment, and $P_{\rm repeat}$ encodes the probability that the retained remnant participates in a subsequent merger.
For $P_{\rm repeat}$, we adopt the following simple phenomenological prescription:
\begin{equation}
P_{\rm repeat} = \epsilon_E \exp \left(-\frac{t_{\rm DF,return}}{\tau_E}\right),
\end{equation}
where $t_{\rm DF,return}$ is the dynamical-friction return time of the remnant, $\epsilon_E$ is the maximum efficiency for forming a new merging binary in a given environment and undergoing a subsequent merger, and $\tau_E$ is a characteristic timescale that parametrizes how rapidly hierarchical merger opportunities are suppressed for remnants that remain displaced from the cluster core. Remnants that quickly re-center ($t_{\rm DF,return} \ll \tau_E$) retain a probability $\sim \epsilon_E$ of undergoing a second merger, whereas remnants with long return times are exponentially suppressed.

We evaluate $P_{\rm hier}$ for representative GCs and NSCs using the dynamical-friction return times computed in Section~\ref{sec:fallback}. For globular clusters we adopt fiducial values $\epsilon_{\rm GC}=0.2$ and $\tau_{\rm GC}=0.3\,{\rm Gyr}$, motivated by dynamical simulations of black-hole interactions in dense stellar systems~\cite{Rodriguez:2018pss,Rodriguez:2016vmx,Rodriguez:2017pec}. For nuclear star clusters we adopt $\epsilon_{\rm NSC}=0.4$ and $\tau_{\rm NSC}=0.5\,{\rm Gyr}$, reflecting their higher densities and escape velocities~\cite{Rodriguez:2018pss,Rodriguez:2016vmx,Rodriguez:2017pec}. We find that the hierarchical-merger probability is strongly suppressed in GCs (typically at the percent level) for the majority of GW events (Fig.~\ref{fig:gc_nsc_hierarchical_probability}). This suppression arises from two independent effects: (i) the relatively small escape velocities of globular clusters, which limit the retention probability of merger remnants, and (ii) large recoil-induced displacements that lead to long dynamical-friction return times. In contrast, NSCs exhibit significantly higher hierarchical-merger probabilities ($\sim 3\%$--$15\%$) due to their larger escape velocities and shorter dynamical-friction timescales. These results suggest that hierarchical mergers among the currently observed GW events are unlikely to occur in typical GCs but may be more common in NSCs or other environments with large escape velocities, such as AGN disks.

\begin{figure}
    \centering
    \includegraphics[width=\columnwidth]{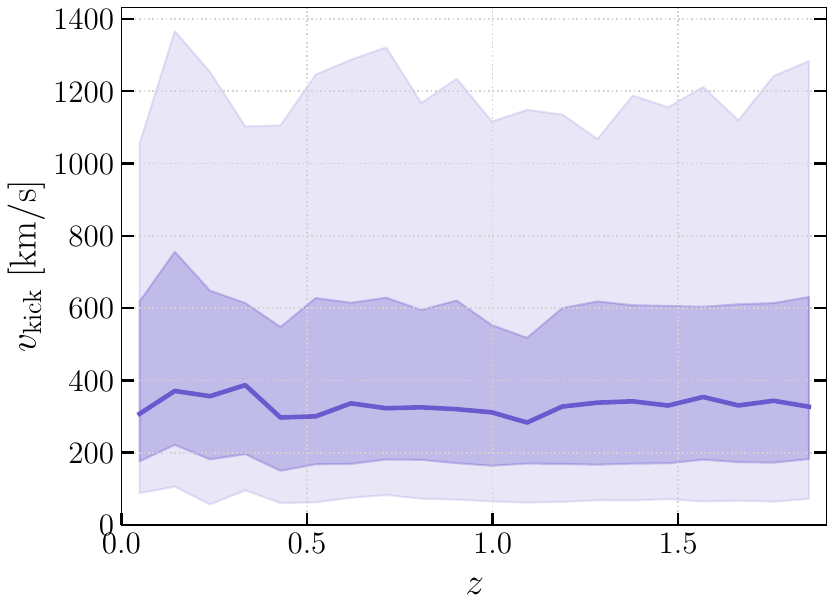}
    \caption{\textit{Population recoil kicks as a function of redshift}. We show the median recoil kick velocity inferred from the GWTC--5 VAR--1 population model as a function of redshift. The dark and light shaded regions indicate the 50\% and 90\% credible intervals, respectively. More details are in Section~\ref{sec:astro_pop_redshift}.}
    \label{fig:pvkick_vs_z}
\end{figure}

\subsection{Astrophysical implications of population recoil kicks}
\label{sec:astro_population}

\subsubsection{Redshift evolution of remnant kicks}
\label{sec:astro_pop_redshift}
We now turn to the astrophysical implications of the recoil-kick distributions inferred from the population properties of BBH mergers.
First, we investigate the redshift evolution of the inferred recoil-kick population. In Fig.~\ref{fig:pvkick_vs_z}, we show the median recoil kick velocity, together with the corresponding 50\% and 90\% credible intervals, inferred from the GWTC--5 VAR--1 population model as a function of redshift. We do not find strong evidence for redshift evolution in the inferred recoil-kick population over the redshift range probed by current GW observations.

\subsubsection{Retention probability of GWTC remnant black holes}
\label{sec:astro_pop_pret}
\begin{figure}
    \includegraphics[width=\columnwidth]{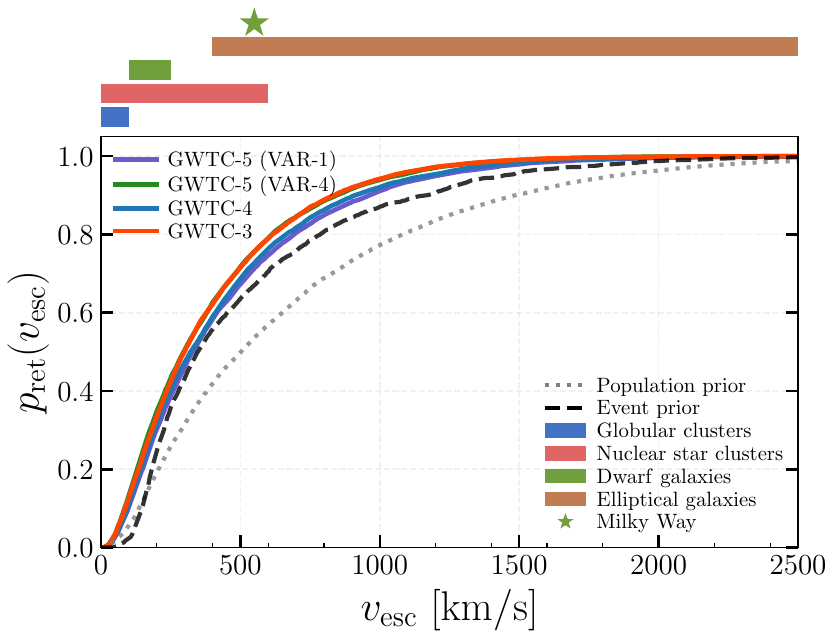}
    \caption{\textit{Retention probability of the inferred BBH population}. We show the retention probability of BBH merger remnants inferred from the GWTC--5 VAR--1 (violet), GWTC--5 VAR--4 (green), GWTC--4 (blue), and GWTC--3 (orange-red) population models as a function of the escape velocity in the range $0$--$2500\,\mathrm{km\,s^{-1}}$. For reference, we indicate representative escape velocity ranges for globular clusters, nuclear star clusters, dwarf galaxies, elliptical galaxies, and the Milky Way. We also show the retention probabilities implied by the event prior and population hyper-prior as black dashed and dotted lines, respectively. More details are in Section~\ref{sec:astro_pop_pret}.}
    \label{fig:gwtc_retention_probability_population}
\end{figure}

Next, we compute the retention probability $p_{\rm ret}(v_{\rm esc})$ inferred from the GWTC--5 VAR--1, GWTC--5 VAR--4, GWTC--4, and GWTC--3 population models as a function of the escape velocity in the range $0$--$2500\,\mathrm{km\,s^{-1}}$ (Fig.~\ref{fig:gwtc_retention_probability_population}). We find that the population-level retention probabilities inferred from the different catalogs are similar. Interestingly, the inferred retention curves lie much closer to those implied by the event prior than to those implied by the population hyper-prior, reflecting the preference of the observed BBH population for relatively modest recoil velocities.

\begin{figure}
    \centering
    \includegraphics[width=\columnwidth]{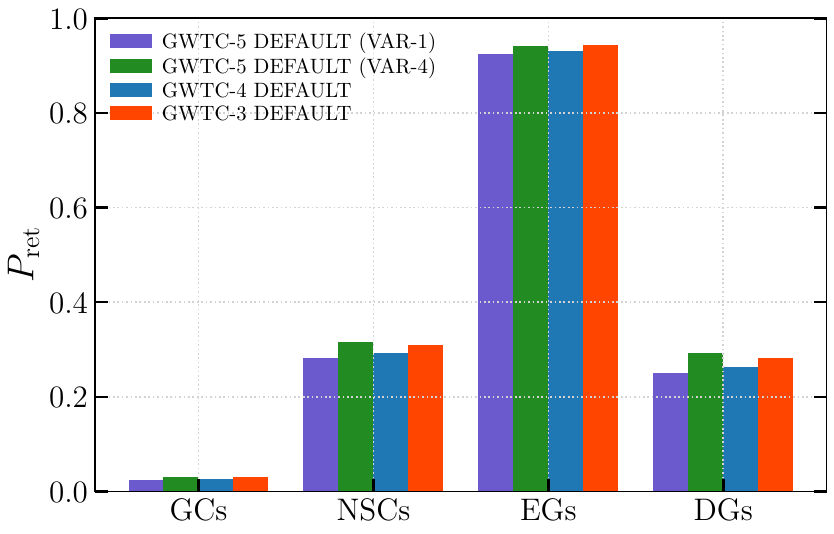}
    \caption{\textit{Retention probability of the inferred BBH population in different host environments}. We show the overall retention probability of BBH merger remnants inferred from the GWTC--5 VAR--1 (violet), GWTC--5 VAR--4 (green), GWTC--4 (blue), and GWTC--3 (orange-red) population models for representative GCs, NSCs, EGs, and DGs. The inferred retention probabilities are broadly consistent across population models. More details are in Section~\ref{sec:astro_pop_pret}.}
    \label{fig:environment_retention_pop}
\end{figure}

We further compute the cumulative retention probability $P_{\rm ret}$ for four representative host environments: GCs, NSCs, DGs, and EGs. The results are shown in Fig.~\ref{fig:environment_retention_pop}. We find that the inferred retention probabilities are highly consistent across all population models. Typical retention probabilities are $\sim0.02$--$0.03$ for GCs, $\sim0.28$--$0.32$ for NSCs, $\sim0.25$--$0.29$ for DGs, and $\sim0.92$--$0.94$ for EGs. These results indicate that BBH merger remnants are rarely retained in globular clusters, are retained only moderately efficiently in nuclear star clusters and dwarf galaxies, and are almost always retained in massive elliptical galaxies. Furthermore, these population-level cumulative retention probabilities are in good agreement with the retention probabilities inferred from individual GW-event recoil-kick posteriors in Section~\ref{sec:astro_events}.

\subsubsection{Spatially displaced GWTC remnant black holes in globular clusters}
\label{sec:astro_pop_gc_displacement}

\begin{figure}
    \centering
    \includegraphics[width=\columnwidth]{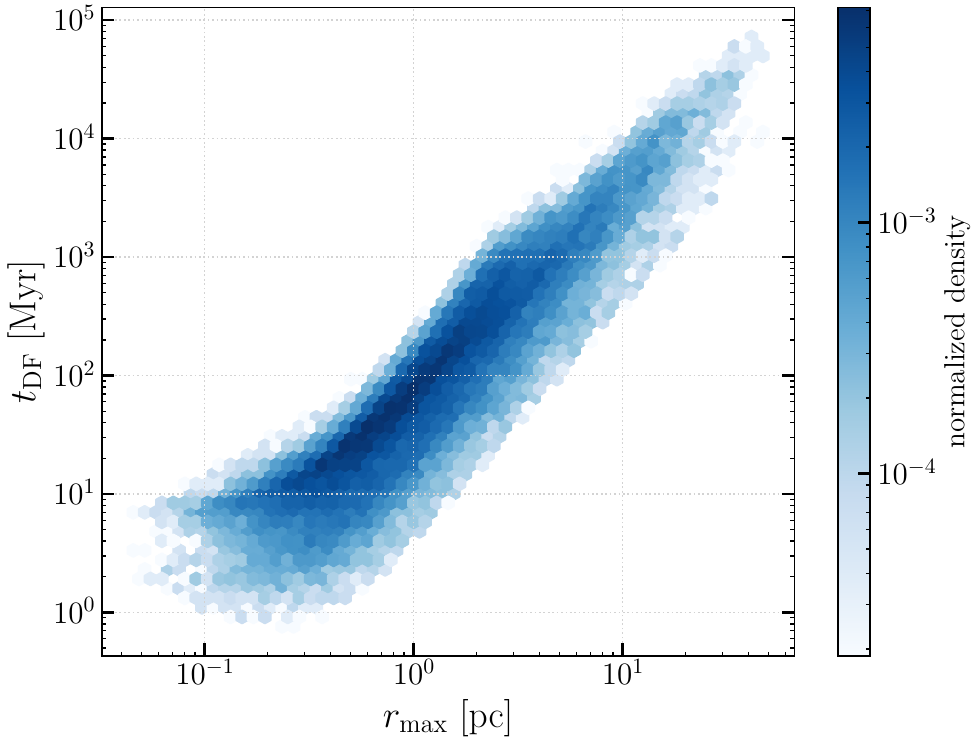}
    \caption{\textit{Maximum recoil displacement and dynamical-friction return time for retained remnants}. We show the joint distribution of the maximum recoil displacement $r_{\rm max}$ and the dynamical-friction return time $t_{\rm DF}$ for retained BBH merger remnants in globular clusters, marginalized over the GWTC--5 VAR--1 population model and the globular-cluster population. The color scale indicates the normalized density. More details are in Section~\ref{sec:astro_pop_gc_displacement}.}
    \label{fig:gc_tdf_rmax_pop}
\end{figure}

While the retention probability quantifies whether a merger remnant remains gravitationally bound to its host cluster, retention alone does not guarantee efficient participation in hierarchical mergers. In Section~\ref{sec:fallback}, we demonstrated this effect for individual GW events by computing the maximum recoil displacement $r_{\rm max}$ and dynamical-friction return time $t_{\rm DF}$ of retained remnants. 
We now quantify the same effect at the population level. We use recoil-kick velocities computed from BBH merger remnants sampled from the GWTC-5 VAR-1 population model. We model the globular-cluster population as Plummer spheres with masses $M_{\rm cl}$ drawn from a log-uniform distribution over $10^4$--$10^6\,M_\odot$ and half-mass radii $r_{\rm h}$ drawn from a log-uniform distribution over $0.5$--$50\,\mathrm{pc}$~\cite{2018MNRAS.478.1520B}. 
The resulting joint distribution of $r_{\rm max}$ and $t_{\rm DF}$ is shown in Fig.~\ref{fig:gc_tdf_rmax_pop}. We find a strong correlation between $r_{\rm max}$ and $t_{\rm DF}$. Remnants displaced by only a fraction of a parsec typically return to the cluster centre on timescales of a few Myr, whereas remnants displaced by several parsecs can have return times of $10^3$--$10^4$ $\mathrm{Myr}$. Consequently, even retained remnants may spend a substantial fraction of a cluster's lifetime outside the dense central regions where dynamical interactions are most efficient.

\begin{figure}
    \centering
    \includegraphics[width=0.9\columnwidth]{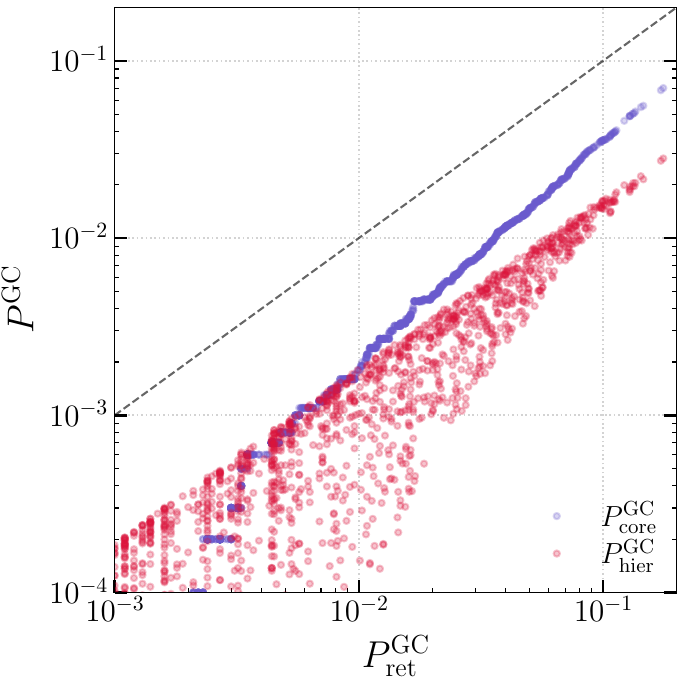}
    \caption{\textit{Retention, core retention, and hierarchical-merger probabilities in globular clusters}. We compare the overall globular-cluster retention probability, $P_{\rm ret}^{\rm GC}$, with the probability that a retained remnant remains confined to the cluster core, $P_{\rm core}^{\rm GC}$ (violet), and the probability that it subsequently participates in a hierarchical merger, $P_{\rm hier}^{\rm GC}$ (red). Each point corresponds to a realization of the globular-cluster population drawn from the GWTC--5 VAR--1 recoil-kick distribution. The dashed line indicates equality with the overall retention probability, $P_{\rm ret}^{\rm GC}$. We find that both $P_{\rm core}^{\rm GC}$ and $P_{\rm hier}^{\rm GC}$ are systematically smaller than $P_{\rm ret}^{\rm GC}$, demonstrating that retention alone substantially overestimates the fraction of remnants that remain dynamically relevant for hierarchical assembly. More details are in Section~\ref{sec:astro_pop_gc_displacement}.}
    \label{fig:gc_pret_pcore_pop}
\end{figure}

Motivated by these results, we define the immediate core-occupancy probability
\begin{equation}
P_{\rm core}^{\rm GC}=\int p(v_{\rm kick})\left[\int_{v_{\rm kick}}^\infty p(v_c)\,dv_c\right]
dv_{\rm kick},
\end{equation}
where \(v_c\) is computed separately for each cluster realization. For a Plummer model with total cluster mass \(M_{\rm cl}\) and scale radius \(a\), we define the characteristic core radius \(r_c\) to be the Plummer core radius,
\begin{equation}
r_c = a\sqrt{2^{2/5}-1},
\end{equation}
and compute \(v_c\) as the recoil speed for which the remnant reaches \(r_c\) at apocenter:
\begin{equation}
v_c=\left[2G M_{\rm cl}\left(\frac{1}{a}-\frac{1}{\sqrt{r_c^2+a^2}}\right)\right]^{1/2}.
\end{equation}
The probability \(P_{\rm core}^{\rm GC}\) is then obtained by averaging the condition \(v_{\rm kick}<v_c\) over both the recoil-kick distribution and the cluster population.
This quantity measures the probability that a merger remnant remains confined to the cluster core immediately after the recoil kick. We compare $P_{\rm core}^{\rm GC}$ with the overall globular-cluster retention probability $P_{\rm ret}^{\rm GC}$ and the hierarchical-merger probability $P_{\rm hier}^{\rm GC}$ in Fig.~\ref{fig:gc_pret_pcore_pop}.
We find that $P_{\rm core}^{\rm GC}$ is systematically smaller than $P_{\rm ret}^{\rm GC}$, typically by about an order of magnitude. This demonstrates that retention alone substantially overestimates the fraction of remnants that remain dynamically relevant for hierarchical assembly. The hierarchical-merger probability $P_{\rm hier}^{\rm GC}$ further accounts for remnants that are initially displaced from the core but subsequently return through dynamical friction and may participate in later mergers. As a result, $P_{\rm hier}^{\rm GC}$ is generally smaller than $P_{\rm core}^{\rm GC}$, although for a small subset of cluster realizations it can be marginally larger due to the contribution from remnants that re-enter the core after a recoil-induced excursion.

These results suggest that, for globular clusters, the prospects for hierarchical mergers are controlled not only by retention but also by the efficiency with which retained remnants return to the cluster core following a recoil kick.

\subsubsection{Spatially displaced GWTC remnant black holes in nuclear star clusters}
\label{sec:astro_pop_nsc_displacement}
\begin{figure}
    \centering
    \includegraphics[width=0.45\textwidth]{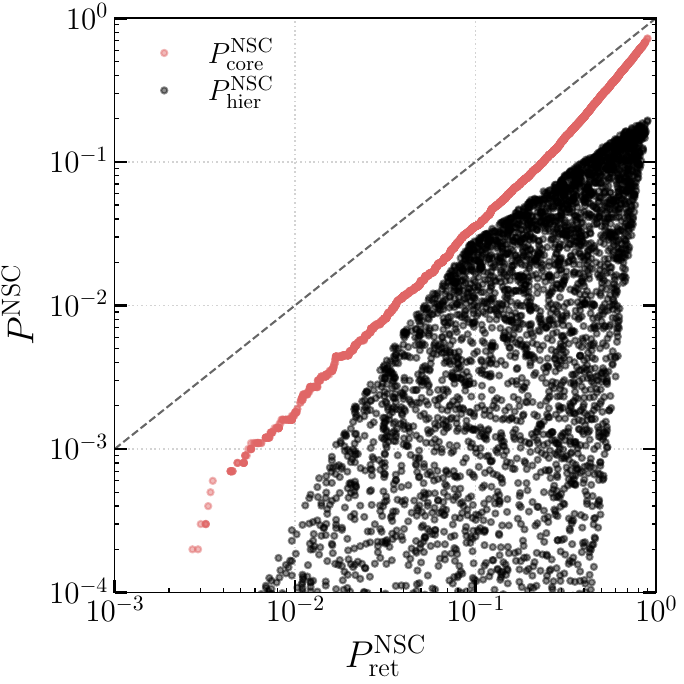}
    \caption{Comparison between the retention probability, $P_{\rm ret}^{\rm NSC}$, the core-retention probability, $P_{\rm core}^{\rm NSC}$, and the hierarchical-merger probability, $P_{\rm hier}^{\rm NSC}$, for nuclear star clusters. The dashed line indicates
    $P^{\rm NSC}=P_{\rm ret}^{\rm NSC}$. While $P_{\rm core}^{\rm NSC}$ closely follows the retention probability, the hierarchical-merger probability is systematically lower, demonstrating that retention alone can substantially overestimate the efficiency of hierarchical mergers in NSCs. More details are in Section~\ref{sec:astro_pop_nsc_displacement}.}
    \label{fig:nsc_pcore_phier}
\end{figure}

We now extend our attention to nuclear star clusters. We sample NSC masses from a log-uniform distribution in the range $M_{\rm cl}\in[10^6,10^8]\,M_\odot$ and half-mass radii from a log-uniform distribution in the range $r_{\rm h}\in[1,20]\,{\rm pc}$~\cite{2020A&ARv..28....4N}. For simplicity, we do not assume the presence of a central massive black hole. For each merger remnant, we compute the recoil velocity and determine whether the remnant remains gravitationally bound to the NSC potential. This defines the retention probability $P_{\rm ret}^{\rm NSC}$. In addition, we define a core-retention probability $P_{\rm core}^{\rm NSC}$, corresponding to the probability that the remnant remains confined to the central region of the NSC following the recoil kick. Figure~\ref{fig:nsc_pcore_phier} shows that $P_{\rm core}^{\rm NSC}$ broadly tracks $P_{\rm ret}^{\rm NSC}$, although deviations from the one-to-one relation are present for systems whose recoil velocities are sufficiently large to displace remnants from the nucleus while still remaining bound to the overall cluster potential.

To quantify the likelihood of hierarchical growth, we also compute $P_{\rm hier}^{\rm NSC}$. As noted in Section~\ref{sec:hierarchical}, we adopt $\epsilon=0.4$ and $\tau=0.5\,{\rm Gyr}$ when evaluating this criterion. For each retained remnant, we estimate the maximum excursion radius $r_{\max}$ induced by the recoil kick and compute the corresponding dynamical-friction timescale $t_{\rm DF}$. We find that $r_{\max}$ ranges from $0.01\,{\rm pc}$ to $99.5\,{\rm pc}$, while the median value of $t_{\rm DF}$ is $1164\,{\rm Myr}$. As expected, $P_{\rm hier}^{\rm NSC}$ is systematically smaller than both $P_{\rm ret}^{\rm NSC}$ and $P_{\rm core}^{\rm NSC}$, often by more than an order of magnitude. In particular, we find that the median $P_{\rm core}^{\rm NSC}$ is approximately $8\%$, while the median $P_{\rm hier}^{\rm NSC}$ is approximately $0.6\%$. This demonstrates that retention alone can significantly overestimate the efficiency of hierarchical mergers in NSCs.

\section{Concluding remarks}
In this work, we inferred recoil-kick velocities for BBH mergers reported in GWTC--1 through GWTC--5, together with LVK candidate IMBH mergers. Our analysis combines the \NRSur{} and \HLZ{} recoil prescriptions and consistently accounts for the evolution of the binary spins to the reference frame required by each model. The results are publicly available at \href{https://github.com/tousifislam/GWTCKick}{https://github.com/tousifislam/GWTCKick}.

At the population level, we infer recoil-kick distributions using the BBH population models released for GWTC--3, GWTC--4, and GWTC--5. Despite differences in the underlying population prescriptions, the inferred recoil distributions are remarkably consistent across catalogs, yielding median recoil velocities of approximately $300$--$330\,\mathrm{km\,s^{-1}}$. Compared to the corresponding hyper-priors, the inferred BBH populations favor systematically lower recoil velocities.

Using both event-level and population-level recoil estimates, we investigated the retention of merger remnants in a variety of astrophysical environments. We find typical retention probabilities of $\sim2$--$3\%$ for globular clusters, $\sim28$--$32\%$ for nuclear star clusters, $\sim25$--$29\%$ for dwarf galaxies, and $\sim92$--$94\%$ for elliptical galaxies. These population-level retention probabilities are broadly consistent with those inferred directly from individual GW events.

Finally, we showed that retention alone is insufficient to assess the prospects for hierarchical mergers. By modeling recoil-induced displacements and dynamical-friction return times in globular clusters and nuclear star clusters, we find that many retained remnants are displaced far from the dense cluster core and require Gyr timescales to return. Consequently, the probability that a remnant remains in, or subsequently returns to, the cluster core is substantially smaller than the overall retention probability. Our results therefore suggest that the efficiency of hierarchical mergers is controlled not only by remnant retention, but also by the dynamical re-centering of retained remnants following a recoil kick.

As future GW catalogs increase both the number and precision of BBH observations, population-level recoil measurements will become increasingly informative. Combined with improved recoil models and astrophysical population studies, they will provide a powerful probe of black-hole assembly, remnant retention, and hierarchical mergers across a wide range of astrophysical environments.

\begin{acknowledgments}
T.I. thanks Jay Wadekar and Tejaswi Venumadhav for helpful discussions, and thanks the anonymous referee for constructive comments and suggestions that helped improve the manuscript.
T.I. is supported in part by the National Science Foundation under Grant No. NSF PHY-2309135 and the Gordon and Betty Moore Foundation Grant No. GBMF7392. 
Use was made of computational facilities purchased with funds from the National Science Foundation (CNS-1725797) and administered by the Center for Scientific Computing (CSC). 
\end{acknowledgments}

\bibliography{kick_references}

@article{CalderonBustillo:2022ldv,
    author = "Calder\'on Bustillo, Juan and Leong, Samson H. W. and Chandra, Koustav",
    title = "{GW190412: measuring a black-hole recoil direction through higher-order gravitational-wave modes}",
    eprint = "2211.03465",
    archivePrefix = "arXiv",
    primaryClass = "gr-qc",
    reportNumber = "LIGO DCC - P2200332",
    month = "11",
    year = "2022"
}

@article{Campanelli:2007cga,
    author = "Campanelli, Manuela and Lousto, Carlos O. and Zlochower, Yosef and Merritt, David",
    title = "{Maximum gravitational recoil}",
    eprint = "gr-qc/0702133",
    archivePrefix = "arXiv",
    doi = "10.1103/PhysRevLett.98.231102",
    journal = "Phys. Rev. Lett.",
    volume = "98",
    pages = "231102",
    year = "2007"
}

@article{Gonzalez:2006md,
    author = "Gonzalez, Jose A. and Sperhake, Ulrich and Bruegmann, Bernd and Hannam, Mark and Husa, Sascha",
    title = "{Total recoil: The Maximum kick from nonspinning black-hole binary inspiral}",
    eprint = "gr-qc/0610154",
    archivePrefix = "arXiv",
    doi = "10.1103/PhysRevLett.98.091101",
    journal = "Phys. Rev. Lett.",
    volume = "98",
    pages = "091101",
    year = "2007"
}

@article{Gonzalez:2007hi,
    author = "Gonzalez, J. A. and Hannam, M. D. and Sperhake, U. and Bruegmann, Bernd and Husa, S.",
    title = "{Supermassive recoil velocities for binary black-hole mergers with antialigned spins}",
    eprint = "gr-qc/0702052",
    archivePrefix = "arXiv",
    doi = "10.1103/PhysRevLett.98.231101",
    journal = "Phys. Rev. Lett.",
    volume = "98",
    pages = "231101",
    year = "2007"
}

@article{Hofmann:2016yih,
    author = "Hofmann, Fabian and Barausse, Enrico and Rezzolla, Luciano",
    title = "{The final spin from binary black holes in quasi-circular orbits}",
    eprint = "1605.01938",
    archivePrefix = "arXiv",
    primaryClass = "gr-qc",
    doi = "10.3847/2041-8205/825/2/L19",
    journal = "Astrophys. J. Lett.",
    volume = "825",
    number = "2",
    pages = "L19",
    year = "2016"
}

@article{Komossa:2008as,
    author = "Komossa, S. and Merritt, David",
    title = "{Gravitational Wave Recoil Oscillations of Black Holes: Implications for Unified Models of Active Galactic Nuclei}",
    eprint = "0811.1037",
    archivePrefix = "arXiv",
    primaryClass = "astro-ph",
    doi = "10.1086/595883",
    journal = "Astrophys. J. Lett.",
    volume = "689",
    pages = "L89",
    year = "2008"
}

@article{LIGOScientific:2018mvr,
    author = "Abbott, B. P. and others",
    collaboration = "LIGO Scientific, Virgo",
    title = "{GWTC-1: A Gravitational-Wave Transient Catalog of Compact Binary Mergers Observed by LIGO and Virgo during the First and Second Observing Runs}",
    eprint = "1811.12907",
    archivePrefix = "arXiv",
    primaryClass = "astro-ph.HE",
    reportNumber = "LIGO-P1800307",
    doi = "10.1103/PhysRevX.9.031040",
    journal = "Phys. Rev. X",
    volume = "9",
    number = "3",
    pages = "031040",
    year = "2019"
}

@article{KAGRA:2021vkt,
    author = "Abbott, R. and others",
    collaboration = "KAGRA, VIRGO, LIGO Scientific",
    title = "{GWTC-3: Compact Binary Coalescences Observed by LIGO and Virgo during the Second Part of the Third Observing Run}",
    eprint = "2111.03606",
    archivePrefix = "arXiv",
    primaryClass = "gr-qc",
    reportNumber = "LIGO-P2000318",
    doi = "10.1103/PhysRevX.13.041039",
    journal = "Phys. Rev. X",
    volume = "13",
    number = "4",
    pages = "041039",
    year = "2023"
}

@article{LIGOScientific:2020ibl,
    author = "Abbott, R. and others",
    collaboration = "LIGO Scientific, Virgo",
    title = "{GWTC-2: Compact Binary Coalescences Observed by LIGO and Virgo During the First Half of the Third Observing Run}",
    eprint = "2010.14527",
    archivePrefix = "arXiv",
    primaryClass = "gr-qc",
    reportNumber = "P2000061",
    doi = "10.1103/PhysRevX.11.021053",
    journal = "Phys. Rev. X",
    volume = "11",
    pages = "021053",
    year = "2021"
}

@article{LIGOScientific:2021usb,
    author = "Abbott, R. and others",
    collaboration = "LIGO Scientific, VIRGO",
    title = "{GWTC-2.1: Deep Extended Catalog of Compact Binary Coalescences Observed by LIGO and Virgo During the First Half of the Third Observing Run}",
    eprint = "2108.01045",
    archivePrefix = "arXiv",
    primaryClass = "gr-qc",
    reportNumber = "LIGO-P2100063",
    month = "8",
    year = "2021"
}

@article{Lousto:2011kp,
    author = "Lousto, Carlos O. and Zlochower, Yosef",
    title = "{Hangup Kicks: Still Larger Recoils by Partial Spin/Orbit Alignment of Black-Hole Binaries}",
    eprint = "1108.2009",
    archivePrefix = "arXiv",
    primaryClass = "gr-qc",
    doi = "10.1103/PhysRevLett.107.231102",
    journal = "Phys. Rev. Lett.",
    volume = "107",
    pages = "231102",
    year = "2011"
}

@book{Maggiore:2007ulw,
    author = "Maggiore, Michele",
    title = "{Gravitational Waves. Vol. 1: Theory and Experiments}",
    doi = "10.1093/acprof:oso/9780198570745.001.0001",
    isbn = "978-0-19-171766-6, 978-0-19-852074-0",
    publisher = "Oxford University Press",
    year = "2007"
}

@book{Maggiore:2018sht,
    author = "Maggiore, Michele",
    title = "{Gravitational Waves. Vol. 2: Astrophysics and Cosmology}",
    isbn = "978-0-19-857089-9",
    publisher = "Oxford University Press",
    month = "3",
    year = "2018"
}

@article{Mahapatra:2021hme,
    author = "Mahapatra, Parthapratim and Gupta, Anuradha and Favata, Marc and Arun, K. G. and Sathyaprakash, B. S.",
    title = "{Remnant Black Hole Kicks and Implications for Hierarchical Mergers}",
    eprint = "2106.07179",
    archivePrefix = "arXiv",
    primaryClass = "astro-ph.HE",
    doi = "10.3847/2041-8213/ac20db",
    journal = "Astrophys. J. Lett.",
    volume = "918",
    pages = "L31",
    year = "2021"
}

@article{Merritt:2004xa,
    author = "Merritt, David and Milosavljevic, Milos and Favata, Marc and Hughes, Scott A. and Holz, Daniel E.",
    title = "{Consequences of gravitational radiation recoil}",
    eprint = "astro-ph/0402057",
    archivePrefix = "arXiv",
    doi = "10.1086/421551",
    journal = "Astrophys. J. Lett.",
    volume = "607",
    pages = "L9--L12",
    year = "2004"
}

@article{Ossokine:2020kjp,
    author = "Ossokine, Serguei and others",
    title = "{Multipolar Effective-One-Body Waveforms for Precessing Binary Black Holes: Construction and Validation}",
    eprint = "2004.09442",
    archivePrefix = "arXiv",
    primaryClass = "gr-qc",
    doi = "10.1103/PhysRevD.102.044055",
    journal = "Phys. Rev. D",
    volume = "102",
    number = "4",
    pages = "044055",
    year = "2020"
}

@article{Pratten:2020ceb,
    author = "Pratten, Geraint and others",
    title = "{Computationally efficient models for the dominant and subdominant harmonic modes of precessing binary black holes}",
    eprint = "2004.06503",
    archivePrefix = "arXiv",
    primaryClass = "gr-qc",
    doi = "10.1103/PhysRevD.103.104056",
    journal = "Phys. Rev. D",
    volume = "103",
    number = "10",
    pages = "104056",
    year = "2021"
}

@article{Ramos-Buades:2023ehm,
    author = "Ramos-Buades, Antoni and Buonanno, Alessandra and Estell\'es, H\'ector and Khalil, Mohammed and Mihaylov, Deyan P. and Ossokine, Serguei and Pompili, Lorenzo and Shiferaw, Mahlet",
    title = "{SEOBNRv5PHM: Next generation of accurate and efficient multipolar precessing-spin effective-one-body waveforms for binary black holes}",
    eprint = "2303.18046",
    archivePrefix = "arXiv",
    primaryClass = "gr-qc",
    month = "3",
    year = "2023"
}

@article{Rodriguez:2016vmx,
    author = "Rodriguez, Carl L. and Zevin, Michael and Pankow, Chris and Kalogera, Vasilliki and Rasio, Frederic A.",
    title = "{Illuminating Black Hole Binary Formation Channels with Spins in Advanced LIGO}",
    eprint = "1609.05916",
    archivePrefix = "arXiv",
    primaryClass = "astro-ph.HE",
    doi = "10.3847/2041-8205/832/1/L2",
    journal = "Astrophys. J. Lett.",
    volume = "832",
    number = "1",
    pages = "L2",
    year = "2016"
}

@article{Rodriguez:2017pec,
    author = "Rodriguez, Carl L. and Amaro-Seoane, Pau and Chatterjee, Sourav and Rasio, Frederic A.",
    title = "{Post-Newtonian Dynamics in Dense Star Clusters: Highly-Eccentric, Highly-Spinning, and Repeated Binary Black Hole Mergers}",
    eprint = "1712.04937",
    archivePrefix = "arXiv",
    primaryClass = "astro-ph.HE",
    doi = "10.1103/PhysRevLett.120.151101",
    journal = "Phys. Rev. Lett.",
    volume = "120",
    number = "15",
    pages = "151101",
    year = "2018"
}

@article{Rodriguez:2018pss,
    author = "Rodriguez, Carl L. and Amaro-Seoane, Pau and Chatterjee, Sourav and Kremer, Kyle and Rasio, Frederic A. and Samsing, Johan and Ye, Claire S. and Zevin, Michael",
    title = "{Post-Newtonian Dynamics in Dense Star Clusters: Formation, Masses, and Merger Rates of Highly-Eccentric Black Hole Binaries}",
    eprint = "1811.04926",
    archivePrefix = "arXiv",
    primaryClass = "astro-ph.HE",
    doi = "10.1103/PhysRevD.98.123005",
    journal = "Phys. Rev. D",
    volume = "98",
    number = "12",
    pages = "123005",
    year = "2018"
}

@article{Romero-Shaw:2020owr,
    author = "Romero-Shaw, I. M. and others",
    title = "{Bayesian inference for compact binary coalescences with bilby: validation and application to the first LIGO\textendash{}Virgo gravitational-wave transient catalogue}",
    eprint = "2006.00714",
    archivePrefix = "arXiv",
    primaryClass = "astro-ph.IM",
    doi = "10.1093/mnras/staa2850",
    journal = "Mon. Not. Roy. Astron. Soc.",
    volume = "499",
    number = "3",
    pages = "3295--3319",
    year = "2020"
}

@article{Varma:2018aht,
    author = "Varma, Vijay and Gerosa, Davide and Stein, Leo C. and H\'ebert, Fran\c{c}ois and Zhang, Hao",
    title = "{High-accuracy mass, spin, and recoil predictions of generic black-hole merger remnants}",
    eprint = "1809.09125",
    archivePrefix = "arXiv",
    primaryClass = "gr-qc",
    doi = "10.1103/PhysRevLett.122.011101",
    journal = "Phys. Rev. Lett.",
    volume = "122",
    number = "1",
    pages = "011101",
    year = "2019"
}

@article{Varma:2020nbm,
    author = "Varma, Vijay and Isi, Maximiliano and Biscoveanu, Sylvia",
    title = "{Extracting the Gravitational Recoil from Black Hole Merger Signals}",
    eprint = "2002.00296",
    archivePrefix = "arXiv",
    primaryClass = "gr-qc",
    doi = "10.1103/PhysRevLett.124.101104",
    journal = "Phys. Rev. Lett.",
    volume = "124",
    number = "10",
    pages = "101104",
    year = "2020"
}

@article{Varma:2022pld,
    author = "Varma, Vijay and Biscoveanu, Sylvia and Islam, Tousif and Shaik, Feroz H. and Haster, Carl-Johan and Isi, Maximiliano and Farr, Will M. and Field, Scott E. and Vitale, Salvatore",
    title = "{Evidence of Large Recoil Velocity from a Black Hole Merger Signal}",
    eprint = "2201.01302",
    archivePrefix = "arXiv",
    primaryClass = "astro-ph.HE",
    doi = "10.1103/PhysRevLett.128.191102",
    journal = "Phys. Rev. Lett.",
    volume = "128",
    number = "19",
    pages = "191102",
    year = "2022"
}

@article{Veitch:2014wba,
    author = "Veitch, J. and others",
    title = "{Parameter estimation for compact binaries with ground-based gravitational-wave observations using the LALInference software library}",
    eprint = "1409.7215",
    archivePrefix = "arXiv",
    primaryClass = "gr-qc",
    reportNumber = "LIGO-P1400152",
    doi = "10.1103/PhysRevD.91.042003",
    journal = "Phys. Rev. D",
    volume = "91",
    number = "4",
    pages = "042003",
    year = "2015"
}

@article{Islam:2023mob,
    author = "Islam, Tousif and Field, Scott E. and Khanna, Gaurav",
    title = "{Remnant black hole properties from numerical-relativity-informed perturbation theory and implications for waveform modeling}",
    eprint = "2301.07215",
    archivePrefix = "arXiv",
    primaryClass = "gr-qc",
    doi = "10.1103/PhysRevD.108.064048",
    journal = "Phys. Rev. D",
    volume = "108",
    number = "6",
    pages = "064048",
    year = "2023"
}

@article{Healy:2022jbh,
    author = "Healy, James and Lousto, Carlos O.",
    title = "{Ultimate Black Hole Recoil: What is the Maximum High-Energy Collision Kick?}",
    eprint = "2301.00018",
    archivePrefix = "arXiv",
    primaryClass = "gr-qc",
    doi = "10.1103/PhysRevLett.131.071401",
    journal = "Phys. Rev. Lett.",
    volume = "131",
    number = "7",
    pages = "071401",
    year = "2023"
}

@article{Zlochower:2015wga,
    author = "Zlochower, Yosef and Lousto, Carlos O.",
    title = "{Modeling the remnant mass, spin, and recoil from unequal-mass, precessing black-hole binaries: The Intermediate Mass Ratio Regime}",
    eprint = "1503.07536",
    archivePrefix = "arXiv",
    primaryClass = "gr-qc",
    doi = "10.1103/PhysRevD.92.024022",
    journal = "Phys. Rev. D",
    volume = "92",
    number = "2",
    pages = "024022",
    year = "2015",
    note = "[Erratum: Phys.Rev.D 94, 029901 (2016)]"
}

@article{Baker:2006vn,
    author = "Baker, John G. and Centrella, Joan and Choi, Dae-Il and Koppitz, Michael and van Meter, James R. and Miller, M. Coleman",
    title = "{Getting a kick out of numerical relativity}",
    eprint = "astro-ph/0603204",
    archivePrefix = "arXiv",
    doi = "10.1086/510448",
    journal = "Astrophys. J. Lett.",
    volume = "653",
    pages = "L93--L96",
    year = "2006"
}

@article{Baker:2007gi,
    author = "Baker, John G. and Boggs, William D. and Centrella, Joan and Kelly, Bernard J. and McWilliams, Sean T. and Miller, M. Coleman and van Meter, James R.",
    title = "{Modeling kicks from the merger of non-precessing black-hole binaries}",
    eprint = "astro-ph/0702390",
    archivePrefix = "arXiv",
    doi = "10.1086/521330",
    journal = "Astrophys. J.",
    volume = "668",
    pages = "1140--1144",
    year = "2007"
}

@article{Baker:2008md,
    author = "Baker, John G. and Boggs, William D. and Centrella, Joan and Kelly, Bernard J. and McWilliams, Sean T. and Miller, M. Coleman and van Meter, James R.",
    title = "{Modeling kicks from the merger of generic black-hole binaries}",
    eprint = "0802.0416",
    archivePrefix = "arXiv",
    primaryClass = "astro-ph",
    doi = "10.1086/590927",
    journal = "Astrophys. J. Lett.",
    volume = "682",
    pages = "L29--L32",
    year = "2008"
}

@article{Bruegmann:2007bri,
    author = "Bruegmann, Bernd and Gonzalez, Jose A. and Hannam, Mark and Husa, Sascha and Sperhake, Ulrich",
    title = "{Exploring black hole superkicks}",
    eprint = "0707.0135",
    archivePrefix = "arXiv",
    primaryClass = "gr-qc",
    doi = "10.1103/PhysRevD.77.124047",
    journal = "Phys. Rev. D",
    volume = "77",
    pages = "124047",
    year = "2008"
}

@article{Campanelli:2007ew,
    author = "Campanelli, Manuela and Lousto, Carlos O. and Zlochower, Yosef and Merritt, David",
    title = "{Large merger recoils and spin flips from generic black-hole binaries}",
    eprint = "gr-qc/0701164",
    archivePrefix = "arXiv",
    doi = "10.1086/516712",
    journal = "Astrophys. J. Lett.",
    volume = "659",
    pages = "L5--L8",
    year = "2007"
}

@article{Choi:2007eu,
    author = "Choi, Dae-Il and Kelly, Bernard J. and Boggs, William D. and Baker, John G. and Centrella, Joan and van Meter, James",
    title = "{Recoiling from a kick in the head-on collision of spinning black holes}",
    eprint = "gr-qc/0702016",
    archivePrefix = "arXiv",
    doi = "10.1103/PhysRevD.76.104026",
    journal = "Phys. Rev. D",
    volume = "76",
    pages = "104026",
    year = "2007"
}

@article{Dain:2008ck,
    author = "Dain, Sergio and Lousto, Carlos O. and Zlochower, Yosef",
    title = "{Extra-Large Remnant Recoil Velocities and Spins from Near-Extremal-Bowen-York-Spin Black-Hole Binaries}",
    eprint = "0803.0351",
    archivePrefix = "arXiv",
    primaryClass = "gr-qc",
    doi = "10.1103/PhysRevD.78.024039",
    journal = "Phys. Rev. D",
    volume = "78",
    pages = "024039",
    year = "2008"
}

@article{Healy:2008js,
    author = "Healy, James and Herrmann, Frank and Hinder, Ian and Shoemaker, Deirdre M. and Laguna, Pablo and Matzner, Richard A.",
    title = "{Superkicks in Hyperbolic Encounters of Binary Black Holes}",
    eprint = "0807.3292",
    archivePrefix = "arXiv",
    primaryClass = "gr-qc",
    doi = "10.1103/PhysRevLett.102.041101",
    journal = "Phys. Rev. Lett.",
    volume = "102",
    pages = "041101",
    year = "2009"
}

@article{Herrmann:2006cd,
    author = "Herrmann, F. and Hinder, I. and Shoemaker, D. and Laguna, P.",
    editor = "Merkowitz, Stephen M. and Livas, Jeffrey C.",
    title = "{Binary black holes and recoil velocities}",
    doi = "10.1063/1.2405025",
    journal = "AIP Conf. Proc.",
    volume = "873",
    number = "1",
    pages = "89--93",
    year = "2006"
}

@article{Lousto:2007db,
    author = "Lousto, Carlos O. and Zlochower, Yosef",
    title = "{Further insight into gravitational recoil}",
    eprint = "0708.4048",
    archivePrefix = "arXiv",
    primaryClass = "gr-qc",
    doi = "10.1103/PhysRevD.77.044028",
    journal = "Phys. Rev. D",
    volume = "77",
    pages = "044028",
    year = "2008"
}

@article{Herrmann:2007ac,
    author = "Herrmann, Frank and Hinder, Ian and Shoemaker, Deirdre and Laguna, Pablo and Matzner, Richard A.",
    title = "{Gravitational recoil from spinning binary black hole mergers}",
    eprint = "gr-qc/0701143",
    archivePrefix = "arXiv",
    doi = "10.1086/513603",
    journal = "Astrophys. J.",
    volume = "661",
    pages = "430--436",
    year = "2007"
}

@article{Herrmann:2007ex,
    author = "Herrmann, Frank and Hinder, Ian and Shoemaker, Deirdre M. and Laguna, Pablo and Matzner, Richard A.",
    title = "{Binary Black Holes: Spin Dynamics and Gravitational Recoil}",
    eprint = "0706.2541",
    archivePrefix = "arXiv",
    primaryClass = "gr-qc",
    doi = "10.1103/PhysRevD.76.084032",
    journal = "Phys. Rev. D",
    volume = "76",
    pages = "084032",
    year = "2007"
}

@article{Herrmann:2007cwl,
    author = "Herrmann, Frank and Hinder, Ian and Shoemaker, Deirdre and Laguna, Pablo",
    title = "{Unequal mass binary black hole plunges and gravitational recoil}",
    doi = "10.1088/0264-9381/24/12/S04",
    journal = "Class. Quant. Grav.",
    volume = "24",
    number = "12",
    pages = "S33--S42",
    year = "2007"
}

@article{Holley-Bockelmann:2007hmm,
    author = "Holley-Bockelmann, Kelly and Gultekin, Kayhan and Shoemaker, Deirdre and Yunes, Nico",
    title = "{Gravitational Wave Recoil and the Retention of Intermediate Mass Black Holes}",
    eprint = "0707.1334",
    archivePrefix = "arXiv",
    primaryClass = "astro-ph",
    doi = "10.1086/591218",
    journal = "Astrophys. J.",
    volume = "686",
    pages = "829",
    year = "2008"
}

@article{Jaramillo:2011re,
    author = "Jaramillo, Jose Luis and Panosso Macedo, Rodrigo and Moesta, Philipp and Rezzolla, Luciano",
    title = "{Black-hole horizons as probes of black-hole dynamics I: post-merger recoil in head-on collisions}",
    eprint = "1108.0060",
    archivePrefix = "arXiv",
    primaryClass = "gr-qc",
    doi = "10.1103/PhysRevD.85.084030",
    journal = "Phys. Rev. D",
    volume = "85",
    pages = "084030",
    year = "2012"
}

@article{Koppitz:2007ev,
    author = "Koppitz, Michael and Pollney, Denis and Reisswig, Christian and Rezzolla, Luciano and Thornburg, Jonathan and Diener, Peter and Schnetter, Erik",
    title = "{Recoil Velocities from Equal-Mass Binary-Black-Hole Mergers}",
    eprint = "gr-qc/0701163",
    archivePrefix = "arXiv",
    doi = "10.1103/PhysRevLett.99.041102",
    journal = "Phys. Rev. Lett.",
    volume = "99",
    pages = "041102",
    year = "2007"
}

@article{Lousto:2008dn,
    author = "Lousto, Carlos O. and Zlochower, Yosef",
    title = "{Modeling gravitational recoil from precessing highly-spinning unequal-mass black-hole binaries}",
    eprint = "0805.0159",
    archivePrefix = "arXiv",
    primaryClass = "gr-qc",
    doi = "10.1103/PhysRevD.79.064018",
    journal = "Phys. Rev. D",
    volume = "79",
    pages = "064018",
    year = "2009"
}

@article{Lousto:2010xk,
    author = "Lousto, Carlos O. and Zlochower, Yosef",
    title = "{Modeling maximum astrophysical gravitational recoil velocities}",
    eprint = "1011.0593",
    archivePrefix = "arXiv",
    primaryClass = "gr-qc",
    doi = "10.1103/PhysRevD.83.024003",
    journal = "Phys. Rev. D",
    volume = "83",
    pages = "024003",
    year = "2011"
}

@article{Schnittman:2007ij,
    author = "Schnittman, Jeremy D. and Buonanno, Alessandra and van Meter, James R. and Baker, John G. and Boggs, William D. and Centrella, Joan and Kelly, Bernard J. and McWilliams, Sean T.",
    title = "{Anatomy of the binary black hole recoil: A multipolar analysis}",
    eprint = "0707.0301",
    archivePrefix = "arXiv",
    primaryClass = "gr-qc",
    doi = "10.1103/PhysRevD.77.044031",
    journal = "Phys. Rev. D",
    volume = "77",
    pages = "044031",
    year = "2008"
}

@article{Pollney:2007ss,
    author = "Pollney, Denis and others",
    title = "{Recoil velocities from equal-mass binary black-hole mergers: A Systematic investigation of spin-orbit aligned configurations}",
    eprint = "0707.2559",
    archivePrefix = "arXiv",
    primaryClass = "gr-qc",
    doi = "10.1103/PhysRevD.76.124002",
    journal = "Phys. Rev. D",
    volume = "76",
    pages = "124002",
    year = "2007"
}

@article{Rezzolla:2010df,
    author = "Rezzolla, Luciano and Macedo, Rodrigo P. and Jaramillo, Jose Luis",
    title = "{Understanding the 'anti-kick' in the merger of binary black holes}",
    eprint = "1003.0873",
    archivePrefix = "arXiv",
    primaryClass = "gr-qc",
    doi = "10.1103/PhysRevLett.104.221101",
    journal = "Phys. Rev. Lett.",
    volume = "104",
    pages = "221101",
    year = "2010"
}

@article{Lousto:2012gt,
    author = "Lousto, Carlos O. and Zlochower, Yosef",
    title = "{Nonlinear Gravitational Recoil from the Mergers of Precessing Black-Hole Binaries}",
    eprint = "1211.7099",
    archivePrefix = "arXiv",
    primaryClass = "gr-qc",
    doi = "10.1103/PhysRevD.87.084027",
    journal = "Phys. Rev. D",
    volume = "87",
    number = "8",
    pages = "084027",
    year = "2013"
}

@article{Lousto:2012su,
    author = "Lousto, Carlos O. and Zlochower, Yosef and Dotti, Massimo and Volonteri, Marta",
    title = "{Gravitational Recoil From Accretion-Aligned Black-Hole Binaries}",
    eprint = "1201.1923",
    archivePrefix = "arXiv",
    primaryClass = "gr-qc",
    doi = "10.1103/PhysRevD.85.084015",
    journal = "Phys. Rev. D",
    volume = "85",
    pages = "084015",
    year = "2012"
}

@article{Miller:2008en,
    author = "Miller, Sarah H. and Matzner, R. A.",
    title = "{Multipole Analysis of Kicks in Collision of Binary Black Holes}",
    eprint = "0807.3028",
    archivePrefix = "arXiv",
    primaryClass = "gr-qc",
    doi = "10.1007/s10714-008-0682-9",
    journal = "Gen. Rel. Grav.",
    volume = "41",
    pages = "525--539",
    year = "2009"
}

@article{Tichy:2007hk,
    author = "Tichy, Wolfgang and Marronetti, Pedro",
    title = "{Binary black hole mergers: Large kicks for generic spin orientations}",
    eprint = "gr-qc/0703075",
    archivePrefix = "arXiv",
    doi = "10.1103/PhysRevD.76.061502",
    journal = "Phys. Rev. D",
    volume = "76",
    pages = "061502",
    year = "2007"
}

@article{vanMeter:2010md,
    author = "van Meter, James R. and Miller, M. Coleman and Baker, John G. and Boggs, William D. and Kelly, Bernard J.",
    title = "{A General Formula for Black Hole Gravitational Wave Kicks}",
    eprint = "1003.3865",
    archivePrefix = "arXiv",
    primaryClass = "astro-ph.HE",
    doi = "10.1088/0004-637X/719/2/1427",
    journal = "Astrophys. J.",
    volume = "719",
    pages = "1427",
    year = "2010"
}

@article{Zlochower:2010sn,
    author = "Zlochower, Yosef and Campanelli, Manuela and Lousto, Carlos O.",
    editor = "Marolf, Donald and Sudarsky, Daniel",
    title = "{Modeling Gravitational Recoil Using Numerical Relativity}",
    eprint = "1011.2210",
    archivePrefix = "arXiv",
    primaryClass = "gr-qc",
    doi = "10.1088/0264-9381/28/11/114015",
    journal = "Class. Quant. Grav.",
    volume = "28",
    pages = "114015",
    year = "2011"
}

@article{Healy:2014yta,
    author = "Healy, James and Lousto, Carlos O. and Zlochower, Yosef",
    title = "{Remnant mass, spin, and recoil from spin aligned black-hole binaries}",
    eprint = "1406.7295",
    archivePrefix = "arXiv",
    primaryClass = "gr-qc",
    doi = "10.1103/PhysRevD.90.104004",
    journal = "Phys. Rev. D",
    volume = "90",
    number = "10",
    pages = "104004",
    year = "2014"
}

@article{Lousto:2009mf,
    author = "Lousto, Carlos O. and Campanelli, Manuela and Zlochower, Yosef and Nakano, Hiroyuki",
    editor = "Husa, Sascha and Krishnan, Badri",
    title = "{Remnant Masses, Spins and Recoils from the Merger of Generic Black-Hole Binaries}",
    eprint = "0904.3541",
    archivePrefix = "arXiv",
    primaryClass = "gr-qc",
    doi = "10.1088/0264-9381/27/11/114006",
    journal = "Class. Quant. Grav.",
    volume = "27",
    pages = "114006",
    year = "2010"
}

@article{Blanchet:2005rj,
    author = "Blanchet, Luc and Qusailah, Moh'd S. S. and Will, Clifford M.",
    title = "{Gravitational recoil of inspiralling black-hole binaries to second post-Newtonian order}",
    eprint = "astro-ph/0507692",
    archivePrefix = "arXiv",
    doi = "10.1086/497332",
    journal = "Astrophys. J.",
    volume = "635",
    pages = "508",
    year = "2005"
}

@article{Sopuerta:2006wj,
    author = "Sopuerta, Carlos F. and Yunes, Nicolas and Laguna, Pablo",
    title = "{Gravitational Recoil from Binary Black Hole Mergers: The Close-Limit Approximation}",
    eprint = "astro-ph/0608600",
    archivePrefix = "arXiv",
    reportNumber = "IGPG-06-8-2",
    doi = "10.1103/PhysRevD.78.049901",
    journal = "Phys. Rev. D",
    volume = "74",
    pages = "124010",
    year = "2006",
    note = "[Erratum: Phys.Rev.D 75, 069903 (2007), Erratum: Phys.Rev.D 78, 049901 (2008)]"
}

@article{Sopuerta:2006et,
    author = "Sopuerta, Carlos F. and Yunes, Nicolas and Laguna, Pablo",
    title = "{Gravitational recoil velocities from eccentric binary black hole mergers}",
    eprint = "astro-ph/0611110",
    archivePrefix = "arXiv",
    doi = "10.1086/512067",
    journal = "Astrophys. J. Lett.",
    volume = "656",
    pages = "L9--L12",
    year = "2007"
}

@article{Favata:2004wz,
    author = "Favata, Marc and Hughes, Scott A. and Holz, Daniel E.",
    title = "{How black holes get their kicks: Gravitational radiation recoil revisited}",
    eprint = "astro-ph/0402056",
    archivePrefix = "arXiv",
    doi = "10.1086/421552",
    journal = "Astrophys. J. Lett.",
    volume = "607",
    pages = "L5--L8",
    year = "2004"
}

@inproceedings{Hughes:2004ck,
    author = "Hughes, Scott A. and Favata, Marc and Holz, Daniel E.",
    title = "{How black holes get their kicks: Radiation recoil in binary black hole mergers}",
    booktitle = "{Conference on Growing Black Holes: Accretion in a Cosmological Context}",
    eprint = "astro-ph/0408492",
    archivePrefix = "arXiv",
    doi = "10.1007/11403913_64",
    month = "8",
    year = "2004"
}

@article{Fitchett:1983qzq,
    author = "Fitchett, M. J.",
    title = "{The influence of gravitational wave momentum losses on the centre of mass motion of a Newtonian binary system}",
    doi = "10.1093/mnras/203.4.1049",
    journal = "Mon. Not. Roy. Astron. Soc.",
    volume = "203",
    number = "4",
    pages = "1049--1062",
    year = "1983"
}

@article{Fitchett:1984qn,
    author = "Fitchett, M. J. and Detweiler, Steven L.",
    title = "{Linear momentum and gravitational-waves - circular orbits around a schwarzschild black-hole}",
    journal = "Mon. Not. Roy. Astron. Soc.",
    volume = "211",
    pages = "933--942",
    year = "1984"
}

@article{Wiseman:1992dv,
    author = "Wiseman, Alan G.",
    title = "{Coalescing binary systems of compact objects to (post)5/2 Newtonian order. 2. Higher order wave forms and radiation recoil}",
    reportNumber = "WUGRAV-92-2",
    doi = "10.1103/PhysRevD.46.1517",
    journal = "Phys. Rev. D",
    volume = "46",
    pages = "1517--1539",
    year = "1992"
}

@article{Kidder:1995zr,
    author = "Kidder, Lawrence E.",
    title = "{Coalescing binary systems of compact objects to postNewtonian 5/2 order. 5. Spin effects}",
    eprint = "gr-qc/9506022",
    archivePrefix = "arXiv",
    reportNumber = "NU-GR-11, WUGRAV-94-6A",
    doi = "10.1103/PhysRevD.52.821",
    journal = "Phys. Rev. D",
    volume = "52",
    pages = "821--847",
    year = "1995"
}

@article{Nakano:2010kv,
    author = "Nakano, Hiroyuki and Campanelli, Manuela and Lousto, Carlos O. and Zlochower, Yosef",
    editor = "Lehner, Luis and Pfeiffer, Harald P. and Poisson, E.",
    title = "{Perturbative effects of spinning black holes with applications to recoil velocities}",
    eprint = "1011.2767",
    archivePrefix = "arXiv",
    primaryClass = "gr-qc",
    doi = "10.1088/0264-9381/28/13/134005",
    journal = "Class. Quant. Grav.",
    volume = "28",
    pages = "134005",
    year = "2011"
}

@article{Sundararajan:2010sr,
    author = "Sundararajan, Pranesh A. and Khanna, Gaurav and Hughes, Scott A.",
    title = "{Binary black hole merger gravitational waves and recoil in the large mass ratio limit}",
    eprint = "1003.0485",
    archivePrefix = "arXiv",
    primaryClass = "gr-qc",
    doi = "10.1103/PhysRevD.81.104009",
    journal = "Phys. Rev. D",
    volume = "81",
    pages = "104009",
    year = "2010"
}

@article{Price:2013paa,
    author = "Price, Richard H. and Khanna, Gaurav and Hughes, Scott A.",
    title = "{Black hole binary inspiral and trajectory dominance}",
    eprint = "1306.1159",
    archivePrefix = "arXiv",
    primaryClass = "gr-qc",
    doi = "10.1103/PhysRevD.88.104004",
    journal = "Phys. Rev. D",
    volume = "88",
    number = "10",
    pages = "104004",
    year = "2013"
}

@article{Price:2011fm,
    author = "Price, Richard H. and Khanna, Gaurav and Hughes, Scott A.",
    title = "{Systematics of black hole binary inspiral kicks and the slowness approximation}",
    eprint = "1104.0387",
    archivePrefix = "arXiv",
    primaryClass = "gr-qc",
    doi = "10.1103/PhysRevD.83.124002",
    journal = "Phys. Rev. D",
    volume = "83",
    pages = "124002",
    year = "2011"
}

@article{Berti:2012zp,
    author = "Berti, Emanuele and Kesden, Michael and Sperhake, Ulrich",
    title = "{Effects of post-Newtonian Spin Alignment on the Distribution of Black-Hole Recoils}",
    eprint = "1203.2920",
    archivePrefix = "arXiv",
    primaryClass = "astro-ph.HE",
    doi = "10.1103/PhysRevD.85.124049",
    journal = "Phys. Rev. D",
    volume = "85",
    pages = "124049",
    year = "2012"
}

@article{Gultekin:2004pm,
    author = "Gultekin, Kayhan and Miller, M. Coleman and Hamilton, Douglas P.",
    title = "{Growth of intermediate - mass black holes in globular clusters}",
    eprint = "astro-ph/0402532",
    archivePrefix = "arXiv",
    doi = "10.1086/424809",
    journal = "Astrophys. J.",
    volume = "616",
    pages = "221--230",
    year = "2004"
}

@article{Gerosa:2016vip,
    author = "Gerosa, Davide and Moore, Christopher J.",
    title = "{Black hole kicks as new gravitational wave observables}",
    eprint = "1606.04226",
    archivePrefix = "arXiv",
    primaryClass = "gr-qc",
    reportNumber = "LIGO-P1600118",
    doi = "10.1103/PhysRevLett.117.011101",
    journal = "Phys. Rev. Lett.",
    volume = "117",
    number = "1",
    pages = "011101",
    year = "2016"
}

@article{Borchers:2025sid,
    author = "Borchers, Angela and Ye, Claire S. and Fishbach, Maya",
    title = "{Gravitational-wave Kicks Impact the Spins of Black Holes from Hierarchical Mergers}",
    eprint = "2503.21278",
    archivePrefix = "arXiv",
    primaryClass = "astro-ph.HE",
    doi = "10.3847/1538-4357/addec6",
    journal = "Astrophys. J.",
    volume = "987",
    number = "2",
    year = "2025"
}

@article{Gerosa:2021hsc,
    author = "Gerosa, Davide and Giacobbo, Nicola and Vecchio, Alberto",
    title = "{High mass but low spin: an exclusion region to rule out hierarchical black-hole mergers as a mechanism to populate the pair-instability mass gap}",
    eprint = "2104.11247",
    archivePrefix = "arXiv",
    primaryClass = "astro-ph.HE",
    doi = "10.3847/1538-4357/ac00bb",
    journal = "Astrophys. J.",
    volume = "915",
    pages = "56",
    year = "2021"
}

@article{Gerosa:2017kvu,
    author = "Gerosa, Davide and Berti, Emanuele",
    title = "{Are merging black holes born from stellar collapse or previous mergers?}",
    eprint = "1703.06223",
    archivePrefix = "arXiv",
    primaryClass = "gr-qc",
    doi = "10.1103/PhysRevD.95.124046",
    journal = "Phys. Rev. D",
    volume = "95",
    number = "12",
    pages = "124046",
    year = "2017"
}

@article{Antonini:2016gqe,
    author = "Antonini, Fabio and Rasio, Frederic A.",
    title = "{Merging black hole binaries in galactic nuclei: implications for advanced-LIGO detections}",
    eprint = "1606.04889",
    archivePrefix = "arXiv",
    primaryClass = "astro-ph.HE",
    doi = "10.3847/0004-637X/831/2/187",
    journal = "Astrophys. J.",
    volume = "831",
    number = "2",
    pages = "187",
    year = "2016"
}

@article{Yang:2019cbr,
    author = "Yang, Yang and others",
    title = "{Hierarchical Black Hole Mergers in Active Galactic Nuclei}",
    eprint = "1906.09281",
    archivePrefix = "arXiv",
    primaryClass = "astro-ph.HE",
    doi = "10.1103/PhysRevLett.123.181101",
    journal = "Phys. Rev. Lett.",
    volume = "123",
    number = "18",
    pages = "181101",
    year = "2019"
}

@article{Varma:2019csw,
    author = "Varma, Vijay and Field, Scott E. and Scheel, Mark A. and Blackman, Jonathan and Gerosa, Davide and Stein, Leo C. and Kidder, Lawrence E. and Pfeiffer, Harald P.",
    title = "{Surrogate models for precessing binary black hole simulations with unequal masses}",
    eprint = "1905.09300",
    archivePrefix = "arXiv",
    primaryClass = "gr-qc",
    doi = "10.1103/PhysRevResearch.1.033015",
    journal = "Phys. Rev. Research.",
    volume = "1",
    pages = "033015",
    year = "2019"
}

@article{Sperhake:2019wwo,
    author = "Sperhake, U. and Rosca-Mead, R. and Gerosa, D. and Berti, E.",
    title = "{Amplification of superkicks in black-hole binaries through orbital eccentricity}",
    eprint = "1910.01598",
    archivePrefix = "arXiv",
    primaryClass = "gr-qc",
    doi = "10.1103/PhysRevD.101.024044",
    journal = "Phys. Rev. D",
    volume = "101",
    number = "2",
    pages = "024044",
    year = "2020"
}

@article{KAGRA:2021duu,
    author = "Abbott, R. and others",
    collaboration = "KAGRA, VIRGO, LIGO Scientific",
    title = "{Population of Merging Compact Binaries Inferred Using Gravitational Waves through GWTC-3}",
    eprint = "2111.03634",
    archivePrefix = "arXiv",
    primaryClass = "astro-ph.HE",
    reportNumber = "LIGO-P2100239 ; Data release: https://zenodo.org/record/5655785, LIGO-P2100239",
    doi = "10.1103/PhysRevX.13.011048",
    journal = "Phys. Rev. X",
    volume = "13",
    number = "1",
    pages = "011048",
    year = "2023"
}

@article{Islam:2025drw,
    author = "Islam, Tousif and Wadekar, Digvijay",
    title = "{Accurate models for recoil velocity distribution in black hole mergers with comparable to extreme mass-ratios and their astrophysical implications}",
    eprint = "2511.11536",
    archivePrefix = "arXiv",
    primaryClass = "gr-qc",
    month = "11",
    year = "2025"
}

@article{Islam:2023zzj,
    author = "Islam, Tousif and Vajpeyi, Avi and Shaik, Feroz H. and Haster, Carl-Johan and Varma, Vijay and Field, Scott E. and Lange, Jacob and O'Shaughnessy, Richard and Smith, Rory",
    title = "{Analysis of GWTC-3 with fully precessing numerical relativity surrogate models}",
    eprint = "2309.14473",
    archivePrefix = "arXiv",
    primaryClass = "gr-qc",
    doi = "10.1103/48ck-2fff",
    journal = "Phys. Rev. D",
    volume = "112",
    number = "4",
    pages = "044001",
    year = "2025"
}

@article{Barausse:2012qz,
    author = "Barausse, Enrico and Morozova, Viktoriya and Rezzolla, Luciano",
    title = "{On the mass radiated by coalescing black-hole binaries}",
    eprint = "1206.3803",
    archivePrefix = "arXiv",
    primaryClass = "gr-qc",
    doi = "10.1088/0004-637X/758/1/63",
    journal = "Astrophys. J.",
    volume = "758",
    pages = "63",
    year = "2012",
    note = "[Erratum: Astrophys.J. 786, 76 (2014)]"
}

@article{LIGOScientific:2025slb,
    author = "Abac, A. G. and others",
    collaboration = "LIGO Scientific, VIRGO, KAGRA",
    title = "{GWTC-4.0: Updating the Gravitational-Wave Transient Catalog with Observations from the First Part of the Fourth LIGO-Virgo-KAGRA Observing Run}",
    eprint = "2508.18082",
    archivePrefix = "arXiv",
    primaryClass = "gr-qc",
    reportNumber = "LIGO-P2400386",
    month = "8",
    year = "2025"
}

@article{Colleoni:2024knd,
    author = "Colleoni, Marta and Vidal, Felip A. Ramis and Garc{\'\i}a-Quir{\'o}s, Cecilio and Ak{\c{c}}ay, Sarp and Bera, Sayantani",
    title = "{Fast frequency-domain gravitational waveforms for precessing binaries with a new twist}",
    eprint = "2412.16721",
    archivePrefix = "arXiv",
    primaryClass = "gr-qc",
    doi = "10.1103/PhysRevD.111.104019",
    journal = "Phys. Rev. D",
    volume = "111",
    number = "10",
    pages = "104019",
    year = "2025"
}

@article{Ruiz-Rocha:2025yno,
    author = "Ruiz-Rocha, Krystal and Yelikar, Anjali B. and Lange, Jacob and Gabella, William and Weller, Robert A. and O'Shaughnessy, Richard and Holley-Bockelmann, Kelly and Jani, Karan",
    title = "{Properties of {\textquotedblleft}Lite{\textquotedblright} Intermediate-mass Black Hole Candidates in LIGO-Virgo{\textquoteright}s Third Observing Run}",
    eprint = "2502.17681",
    archivePrefix = "arXiv",
    primaryClass = "astro-ph.HE",
    doi = "10.3847/2041-8213/adc5f8",
    journal = "Astrophys. J. Lett.",
    volume = "985",
    number = "2",
    pages = "L37",
    year = "2025"
}

@ARTICLE{2018MNRAS.478.1520B,
       author = {{Baumgardt}, H. and {Hilker}, M.},
        title = "{A catalogue of masses, structural parameters, and velocity dispersion profiles of 112 Milky Way globular clusters}",
      journal = {Monthly Notices of the Royal Astronomical Society},
         year = 2018,
        month = aug,
       volume = {478},
       number = {2},
        pages = {1520-1557},
          doi = {10.1093/mnras/sty1057},
archivePrefix = {arXiv},
       eprint = {1804.08359},
 primaryClass = {astro-ph.GA},
       adsurl = {https://ui.adsabs.harvard.edu/abs/2018MNRAS.478.1520B}
}

@ARTICLE{2008ApJ...678..780G,
       author = {{Gualandris}, Alessia and {Merritt}, David},
        title = "{Ejection of Supermassive Black Holes from Galaxy Cores}",
      journal = {Astrophysical Journal},
         year = 2008,
        month = may,
       volume = {678},
       number = {2},
        pages = {780-797},
          doi = {10.1086/586877},
archivePrefix = {arXiv},
       eprint = {0708.0771},
 primaryClass = {astro-ph},
       adsurl = {https://ui.adsabs.harvard.edu/abs/2008ApJ...678..780G}
}

@article{LIGOScientific:2021tfm,
    author = "Abbott, Rich and others",
    collaboration = "LIGO Scientific, VIRGO, KAGRA",
    title = "{Search for intermediate-mass black hole binaries in the third observing run of Advanced LIGO and Advanced Virgo}",
    eprint = "2105.15120",
    archivePrefix = "arXiv",
    primaryClass = "astro-ph.HE",
    reportNumber = "LIGO-P2100025",
    doi = "10.1051/0004-6361/202141452",
    journal = "Astron. Astrophys.",
    volume = "659",
    pages = "A84",
    year = "2022"
}

@article{LIGOScientific:2025brd,
    author = "Abac, A. G. and others",
    collaboration = "LIGO Scientific, Virgo, KAGRA",
    title = "{GW241011 and GW241110: Exploring Binary Formation and Fundamental Physics with Asymmetric, High-spin Black Hole Coalescences}",
    eprint = "2510.26931",
    archivePrefix = "arXiv",
    primaryClass = "astro-ph.HE",
    reportNumber = "LIGO-P2500402",
    doi = "10.3847/2041-8213/ae0d54",
    journal = "Astrophys. J. Lett.",
    volume = "993",
    number = "1",
    pages = "L21",
    year = "2025"
}

@article{LIGOScientific:2025rid,
    author = "Abac, A. G. and others",
    collaboration = "LIGO Scientific, Virgo, KAGRA",
    title = "{GW250114: Testing Hawking{\textquoteright}s Area Law and the Kerr Nature of Black Holes}",
    eprint = "2509.08054",
    archivePrefix = "arXiv",
    primaryClass = "gr-qc",
    reportNumber = "LIGO-P2500421",
    doi = "10.1103/kw5g-d732",
    journal = "Phys. Rev. Lett.",
    volume = "135",
    number = "11",
    pages = "111403",
    year = "2025"
}

@article{Yu:2023lml,
    author = "Yu, Hang and Roulet, Javier and Venumadhav, Tejaswi and Zackay, Barak and Zaldarriaga, Matias",
    title = "{Accurate and efficient waveform model for precessing binary black holes}",
    eprint = "2306.08774",
    archivePrefix = "arXiv",
    primaryClass = "gr-qc",
    doi = "10.1103/PhysRevD.108.064059",
    journal = "Phys. Rev. D",
    volume = "108",
    number = "6",
    pages = "064059",
    year = "2023"
}

@article{MacUilliam:2024oif,
    author = "Mac Uilliam, Jake and Akcay, Sarp and Thompson, Jonathan E.",
    title = "{Survey of four precessing waveform models for binary black hole systems}",
    eprint = "2402.06781",
    archivePrefix = "arXiv",
    primaryClass = "gr-qc",
    doi = "10.1103/PhysRevD.109.084077",
    journal = "Phys. Rev. D",
    volume = "109",
    number = "8",
    pages = "084077",
    year = "2024"
}

@ARTICLE{2018A&A...616L...9M,
       author = {{Monari}, G. and {Famaey}, B. and {Carrillo}, I. and {Piffl}, T. and {Steinmetz}, M. and {Wyse}, R.~F.~G. and {Anders}, F. and {Chiappini}, C. and {Jan{\ss}en}, K.},
        title = "{The escape speed curve of the Galaxy obtained from Gaia DR2 implies a heavy Milky Way}",
      journal = {Astronomy \& Astrophysics},
         year = 2018,
        month = aug,
       volume = {616},
          eid = {L9},
        pages = {L9},
          doi = {10.1051/0004-6361/201833748},
archivePrefix = {arXiv},
       eprint = {1807.04565},
 primaryClass = {astro-ph.GA},
       adsurl = {https://ui.adsabs.harvard.edu/abs/2018A&A...616L...9M}
}

@ARTICLE{2023ApJ...955..116L,
       author = {{Lam}, Casey Y. and {Lu}, Jessica R.},
        title = "{A Reanalysis of the Isolated Black Hole Candidate OGLE-2011-BLG-0462/MOA-2011-BLG-191}",
      journal = {The Astrophysical Journal},
         year = 2023,
        month = oct,
       volume = {955},
       number = {2},
          eid = {116},
        pages = {116},
          doi = {10.3847/1538-4357/aced4a},
archivePrefix = {arXiv},
       eprint = {2308.03302},
 primaryClass = {astro-ph.SR},
       adsurl = {https://ui.adsabs.harvard.edu/abs/2023ApJ...955..116L}
}

@article{Doctor:2021qfn,
    author = "Doctor, Zoheyr and Farr, Ben and Holz, Daniel E.",
    title = "{Black Hole Leftovers: The Remnant Population from Binary Black Hole Mergers}",
    eprint = "2103.04001",
    archivePrefix = "arXiv",
    primaryClass = "astro-ph.HE",
    doi = "10.3847/2041-8213/ac0334",
    journal = "Astrophys. J. Lett.",
    volume = "914",
    number = "1",
    pages = "L18",
    year = "2021"
}

@article{LIGOScientific:2026ctl,
    collaboration = "LIGO Scientific, VIRGO, KAGRA",
    title = "{GWTC-5.0: Population Properties of Merging Compact Binaries}",
    eprint = "2605.27226",
    archivePrefix = "arXiv",
    primaryClass = "astro-ph.HE",
    reportNumber = "LIGO-P2600045",
    month = "5",
    year = "2026"
}

@article{LIGOScientific:2026wfs,
    collaboration = "LIGO Scientific, VIRGO, KAGRA",
    title = "{GWTC-5.0: Observations from the Second Part of the Fourth LIGO-Virgo-KAGRA Observing Run and Updates to the Gravitational-Wave Transient Catalog}",
    eprint = "2605.27225",
    archivePrefix = "arXiv",
    primaryClass = "gr-qc",
    reportNumber = "LIGO-P2600152",
    month = "5",
    year = "2026"
}

@article{LIGOScientific:2025pvj,
    author = "Abac, A. G. and others",
    collaboration = "LIGO Scientific, VIRGO, KAGRA",
    title = "{GWTC-4.0: Population Properties of Merging Compact Binaries}",
    eprint = "2508.18083",
    archivePrefix = "arXiv",
    primaryClass = "astro-ph.HE",
    reportNumber = "LIGO-P2400004",
    month = "8",
    year = "2025"
}

@article{Mahapatra:2024qsy,
    author = "Mahapatra, Parthapratim and Chattopadhyay, Debatri and Gupta, Anuradha and Antonini, Fabio and Favata, Marc and Sathyaprakash, B. S. and Arun, K. G.",
    title = "{Reconstructing the Genealogy of LIGO-Virgo Black Holes}",
    eprint = "2406.06390",
    archivePrefix = "arXiv",
    primaryClass = "astro-ph.HE",
    reportNumber = "LIGO preprint number P2400227",
    doi = "10.3847/1538-4357/ad781b",
    journal = "Astrophys. J.",
    volume = "975",
    number = "1",
    pages = "117",
    year = "2024"
}

@book{BinneyTremaine2008,
  author    = {Binney, James and Tremaine, Scott},
  title     = {Galactic Dynamics},
  edition   = {2},
  year      = {2008},
  publisher = {Princeton University Press},
  address   = {Princeton, NJ}
}

@article{Chandrasekhar1943,
  author  = {Chandrasekhar, S.},
  title   = {Dynamical Friction. I. General Considerations: the Coefficient of Dynamical Friction},
  journal = {Astrophysical Journal},
  volume  = {97},
  pages   = {255--262},
  year    = {1943}
}

@article{Plummer1911,
  author  = {Plummer, H. C.},
  title   = {On the problem of distribution in globular star clusters},
  journal = {Monthly Notices of the Royal Astronomical Society},
  volume  = {71},
  pages   = {460--470},
  year    = {1911}
}

@ARTICLE{2020A&ARv..28....4N,
       author = {{Neumayer}, Nadine and {Seth}, Anil and {B{\"o}ker}, Torsten},
        title = "{Nuclear star clusters}",
      journal = {The Astronomy and Astrophysics Review},
         year = 2020,
        month = jul,
       volume = {28},
       number = {1},
          eid = {4},
        pages = {4},
          doi = {10.1007/s00159-020-00125-0},
archivePrefix = {arXiv},
       eprint = {2001.03626},
 primaryClass = {astro-ph.GA},
       adsurl = {https://ui.adsabs.harvard.edu/abs/2020A&ARv..28....4N}
}

\appendix

\section{Displacement and return times for a synthetic BBH population}
\label{app:displacement}

To provide intuition for the dependence of recoil-induced displacements and dynamical-friction return times on the remnant properties, we construct a synthetic population of $10{,}000$ BBH mergers. We sample with component masses $m_1, m_2 \sim \mathcal{U}[3, 60]\,M_\odot$, dimensionless spin magnitudes $|\chi_1|, |\chi_2| \sim \mathcal{U}[0,1]$, isotropic spin tilts $\theta_{1,2}$, and uniformly distributed azimuthal spin angles $\phi_{1,2}$. Recoil kicks are computed with the combined model used throughout this work: \NRSur{} for $q \geq 1/6$ and the \HLZ{} formula for $q < 1/6$ (secc Section~\ref{sec:model}). Remnant masses are obtained from the fitting formulas provided in Refs.~\cite{Barausse:2012qz,Hofmann:2016yih}.

The host environment is chosen to resemble a compact, massive globular cluster. We model it as a Plummer sphere with $M_{\rm cl} = 10^6\,M_\odot$ and half-mass radius $r_h = 1\,$pc (scale radius $a = r_h / 1.305 \approx 0.77\,$pc), giving a central escape speed $v_{\rm esc} \approx 106\,\mathrm{km\,s^{-1}}$ and a core-crossing threshold $v_c \approx 57\,\mathrm{km\,s^{-1}}$. A remnant is counted as cluster-retained only if its apocentre lies within the tidal truncation radius $r_t = 5\,r_h = 5\,$pc, equivalent to $v_{\rm kick} \leq 0.92\,v_{\rm esc}$.

\begin{figure}
    \centering
    \includegraphics[width=\columnwidth]{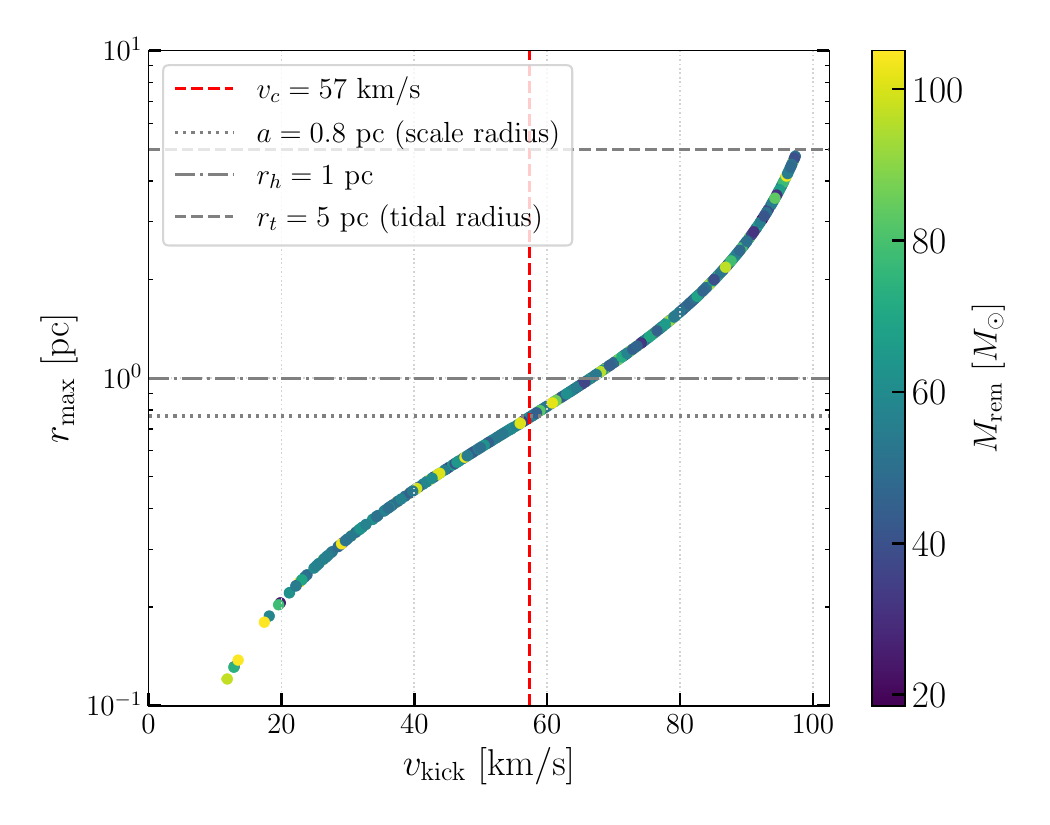}
    \vspace{0.3cm}
    \includegraphics[width=\columnwidth]{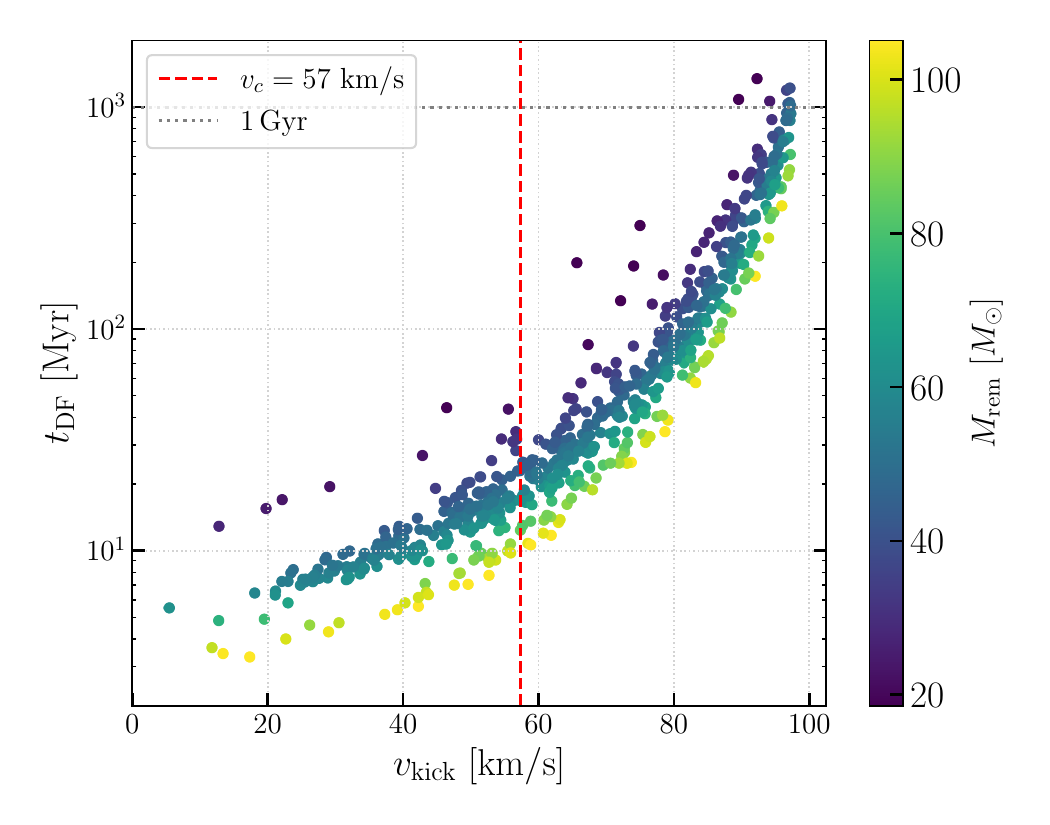}
    \caption{\textit{Dependence of recoil-induced displacements and dynamical-friction return times on recoil kick velocity}.
    \textbf{Top panel:} Apocentre displacement $r_{\max}$ as a function of recoil kick velocity $v_{\rm kick}$ for a synthetic population of retained BBH merger remnants in an idealized Plummer-sphere globular cluster ($M_{\rm cl}=10^6\,M_\odot$, $r_h=1\,{\rm pc}$). The vertical dashed line marks the core-crossing speed $v_c \approx 57\,{\rm km\,s^{-1}}$, while the horizontal lines indicate the scale radius $a$, half-mass radius $r_h$, and tidal radius $r_t$. 
    \textbf{Bottom panel:} Orbit-averaged Chandrasekhar dynamical-friction return time $t_{\rm DF}$ as a function of $v_{\rm kick}$. Points are colored by remnant mass $M_{\rm rem}$. The horizontal dotted line indicates $1\,{\rm Gyr}$. More details are in Appendix~\ref{app:displacement}.}
    \label{fig:appendix_scalings}
\end{figure}

Figure~\ref{fig:appendix_scalings} illustrates how recoil-induced displacements and dynamical-friction return times depend on the recoil kick velocity and remnant mass for a synthetic population of BBH merger remnants. The upper panel shows the apocentre displacement $r_{\max}$ as a function of $v_{\rm kick}$ for all retained remnants. In a Plummer potential, $r_{\max}$ is determined entirely by the ratio $v_{\rm kick}/v_{\rm esc}$ and is therefore independent of the remnant mass, producing the single monotonic relation visible in the figure. Remnants receiving kicks below the core-crossing threshold $v_c$ remain confined within the central region of the cluster, while larger kicks can displace retained remnants beyond the half-mass radius and, for kicks approaching the escape speed, close to the tidal boundary.
The lower panel shows the orbit-averaged Chandrasekhar dynamical-friction return time $t_{\rm DF}$ as a function of $v_{\rm kick}$. Unlike the apocentre displacement, the return time depends on both the kick velocity and the remnant mass. At fixed $v_{\rm kick}$, more massive remnants experience stronger dynamical friction and therefore return to the cluster center more rapidly, whereas lighter remnants remain displaced for longer periods. The return time increases systematically with kick velocity because larger kicks produce more extended orbits that spend a greater fraction of their time in the low-density outer regions of the cluster where dynamical friction is less effective.

For the fiducial cluster considered here, the median return time of retained remnants is $\sim 34\,{\rm Myr}$, and approximately $99\%$ of retained remnants return to the cluster center within $1\,{\rm Gyr}$, well below a typical globular-cluster lifetime. Of the $10{,}000$ synthetic mergers, approximately $6\%$ are retained within the tidal radius and only $\sim 2\%$ remain confined to the cluster core immediately after merger. Although these fractions depend on the adopted prior distribution of binary parameters and should not be interpreted as astrophysical retention fractions, they illustrate a generic feature of recoil dynamics: larger kicks simultaneously increase both the spatial displacement and the re-centering time of merger remnants, thereby reducing the probability that retained remnants participate in hierarchical mergers.

\end{document}